\documentclass[iop,apj]{emulateapj}
\slugcomment{Submitted to The Astrophysical Journal; 1 July 2016; Accepted for publication; 8 December 2016}

\usepackage[]{amsmath}
\usepackage[dvipsnames]{xcolor}
\usepackage{bm}
\usepackage{ulem}
\shorttitle{The GLASS Size--Mass Relation and Low-Mass Passive Galaxies}
\shortauthors{Morishita et al.}

\usepackage{color}			      
\definecolor{midgray}{gray}{0.4}		
\definecolor{orange}{rgb}{1,0.5,0}    
\usepackage{mathptmx}
\usepackage{times}

\usepackage[colorlinks=true, citecolor=NavyBlue, filecolor=NavyBlue, linkcolor=Red, urlcolor=NavyBlue]{hyperref}        
\usepackage{scrextend}
\deffootnote[0.5em]{0em}{1em}{\textsuperscript{\thefootnotemark}\,}

\newcommand{\myemail}{mtaka@astro.ucla.edu}
\newcommand{\simgt}{\,\rlap{\lower 3.5 pt \hbox{$\mathchar \sim$}} \raise
1pt \hbox {$>$}\,}
\newcommand{\simlt}{\,\rlap{\lower 3.5 pt \hbox{$\mathchar \sim$}} \raise
1pt \hbox {$<$}\,}

\newcommand{\logm}{\log M_*/\Msun}
\newcommand{\Msun}{{\rm M}_{\odot}}
\newcommand{\Mstel}{M_{\ast}}

\def\kpc{{\rm kpc}}

\newcommand{\kms}{{\rm km~s^{-1}}}
\newcommand{\hst}{{\it HST}}
\newcommand{\specz}{$z_{\rm spec}$}
\newcommand{\photz}{$z_{\rm phot}$}
\newcommand{\resp}{respectively}

\begin{document}

\title{
The Grism Lens-Amplified Survey from Space (GLASS). IX.  \\
The dual origin of low-mass cluster galaxies as revealed by \\ new structural analyses
}

\author{
Takahiro~Morishita$^{1,2,3}$, 
Louis~E.~Abramson$^{1}$, 
Tommaso~Treu$^{1}$,\\
Benedetta~Vulcani$^{4}$,
Kasper~B.~Schmidt$^{5}$,
Alan~Dressler$^{6}$,
Bianca~M.~Poggianti$^{7}$,
Matthew~A.~Malkan$^{1}$,
Xin~Wang$^{1}$,
Kuang-Han~Huang$^{8}$,
Michele Trenti$^{4}$,
Maru\v{s}a~Brada\v{c}$^{8}$,
and Austin Hoag$^{8}$
}
\affil{$^1$Department of Physics and Astronomy, UCLA, 430 Portola Plaza, Los Angeles, CA 90095-1547, USA; \href{mailto:\myemail}{\myemail}}
\affil{$^2$Astronomical Institute, Tohoku University, Aramaki, Aoba, Sendai 980-8578, Japan}
\affil{$^3$Institute for International Advanced Research and Education, Tohoku University, Aramaki, Aoba, Sendai 980-8578, Japan}
\affil{$^4$School of Physics, The University of Melbourne, VIC 3010, Australia}
\affil{$^5$Leibniz-Institut f\"ur Astrophysik Potsdam (AIP), An der Sternwarte 16, D-14482 Potsdam, Germany}
\affil{$^6$The Observatories of the Carnegie Institution for Science, 813 Santa Barbara Street, Pasadena, CA 91101, USA}
\affil{$^7$INAF-Astronomical Observatory of Padova, Italy}
\affil{$^8$University of California Davis, 1 Shields Avenue, Davis, CA 95616, USA}

\begin{abstract}

Using deep {\it Hubble Frontier Fields} imaging and slitless spectroscopy from the {\it Grism Lens-amplified Survey from Space}, we study 2200 cluster and 1748 field galaxies at $0.2\leq z\leq0.7$ to determine the impact of environment on galaxy size and structure at stellar masses $\logm>7.8$, an unprecedented limit at these redshifts. 
Based on simple assumptions---$r_e= f(\Mstel)$---we find no significant differences in half-light radii ($r_e$) between equal-mass cluster or field systems. 
More complex analyses---$r_e = f(\Mstel, U-V, n, z,\Sigma)$---reveal local density ($\Sigma$) to induce only a 7\%$\pm3\%$ ($95\%$ confidence) reduction in $r_e$ beyond what can be accounted for by $U-V$ color, S\'ersic index ($n$), and redshift ($z$) effects. 
Almost any size difference between galaxies in high- and low-density regions is thus attributable to their different distributions in properties other than environment.
Indeed, we find a clear color--$r_{e}$ correlation in low-mass passive cluster galaxies ($\logm<9.8$) such that bluer systems have larger radii, with the bluest having sizes consistent with equal-mass star-forming galaxies.
We take this as evidence that {\it large-$r_{e}$} low-mass passive cluster galaxies are recently acquired systems that have been environmentally quenched without significant structural transformation (e.g., by ram pressure stripping or starvation). 
Conversely, $\sim20\%$ of {\it small-$r_{e}$} low-mass passive cluster galaxies appear to have been in place since $z\gtrsim3$. 
Given the consistency of the small-$r_{e}$ galaxies' stellar surface densities (and even colors) with those of systems more than ten times as massive, our findings suggest that clusters mark places where galaxy evolution is accelerated for an ancient base population spanning most masses, with late-time additions quenched by environment-specific mechanisms are mainly restricted to the lowest masses.
\end{abstract}



\section{Introduction}
\label{sec:intro}

In terms of a host of properties---color, star formation activity, structure, morphology---clusters harbor different galaxy populations than average (``field") environments \citep[e.g.,][]{hubble31, dressler80}. The mechanisms that produce these differences has been the subject of intense scrutiny. 
While evidence of environmental effects have been seen \citep[e.g.,][]{vollmer09, abramson11, mcpartland16, poggianti16}, their roles and relative importance compared to {\it in situ} galaxy evolution remain poorly understood.
Indeed, the extent to which clusters are agents that halt galaxy evolution, as opposed to {\it tracers} of regions where it has been accelerated, is still under debate (cf.\ \citealt{peng10} with \citealt{dressler80}, \citealt{thomas05}, \citealt{guglielmo15}, \citealt{abramson16}).

One confounding factor is that galaxy-by-galaxy analyses reveal almost no differential environmental effects for systems, e.g., at fixed stellar mass ($\Mstel$) and color \citep{grutzbauch11b}. 
That is, while galaxy {\it populations} are different in low- and high-density regions, representatives of all parts of parameter space seem to exist everywhere \citep[e.g.,][]{dressler13, dressler16, wu14}.\footnote[9]{Excluding the very most- and very least-massive red objects---dwarf ellipticals and cDs/BCGs---which may never exist in isolation \citep{koester07, geha12}.}
This appears to hold even for scaling laws that (seemingly) should contain the signatures of any transformational mechanism, such as the star formation rate--mass and size--mass relations (e.g., \citealt{maltby10, peng10, huertas13, koyama13, allen16}; but see\ \citealt{vulcani10, paccagnella16}, who define environment by spectroscopic membership as opposed to spatial overdensity).

However, a key obstacle to many previous investigations has been their relatively high mass limits of $\logm \gtrsim10$.
In this regime, a system's self-gravity is strong, perhaps protecting it from environmental influences such as ram pressure stripping or harassment \citep[e.g.,][]{dressler83, moore96, treu03, lin14}. 
Furthermore, high-mass galaxies might be subject to internal processes---such as feedback from active galactic nuclei, or the suppression of star formation by morphological structures---that act before they enter the cluster, preventing the latter from having any effect at all \citep[e.g.,][]{martig09, hopkins14}.
To better constrain the physical processes {\it causally related} to environmental density, targeting the low-mass tail of the galaxy population ($\logm\ll10$) is key.

Some studies of the nearest clusters have probed this regime: \citet[][see also \citealt{ferrarese06}]{misgeld11} examined the size--mass relation of Local Group and Coma galaxies at $\logm\gtrsim 6$, and \citet[][]{lisker09} and \citet{toloba15} explored the diversity of dwarf galaxies (dEs, dSphs) in Virgo, with the latter study using kinematical/chemical information to find the evidence of ram pressure stripping-influenced evolution. However, at $z\approx0$, cluster galaxies are so uniformly old that the residual signatures of any transformational mechanisms may be detectable only in the most detailed fossil evidence \citep{mcdermid15}.
Shifting focus to $z\sim0.5$ would alleviate this issue by probing an epoch when clusters were still rapidly assembling, and hence more dramatically reshaping their galaxy populations \citep{butcher78}. The recent advent of ultra-deep multi-band imaging and spectroscopy from space enables such studies of low-mass, mid-$z$ galaxies for the first time.

In this paper, we use data from the {\it Hubble Frontier Fields} \citep[HFF;][]{lotz16} and {\it Grism Lens-Amplified Survey from Space} \citep[GLASS;][]{schmidt14,treu15} to examine the galaxy populations of clusters and the field at $0.2\leq z\leq0.7$ to a hitherto unexplored mass limit of $\logm>7.8$. 

We exploit these data to study the dependence of galaxy size on stellar mass and other structural properties as a function of environmental density using an unprecedented sample of over 3900 cluster and field galaxies. We examine these correlations because: (1) processes that depend on environment---e.g., ram pressure stripping, or mergers (rarer in richer systems)---affect galaxy size and structure in significant and well-defined ways \citep[][and many others]{vandokkum10, damjanov11, newman12, nipoti12, patel13}, and (2) the depth and resolution of new \hst\ imaging enables analyses of galaxy structure that current ground-based observations cannot support. 
This is especially true in the near-infrared, which most directly probes galaxies' stellar mass distributions.

By using a new multidimensional approach that holistically examines galaxies in their natural parameter space---spanning color, size, structure, environmental density, and redshift---our analysis provides a new look at both rapid and long-term environmental influences, yielding a ``4D'' view of the galaxy population at unexplored masses and spatial resolutions. 

We organize our discussion around the central question posed above: how many of the observed differences in cluster/field populations reflect phenomena {\it driven} by clusters versus those {\it traced} by them? 

Ultimately, our results suggest that, while a cluster-specific process similar to ram pressure stripping is indeed operational now, $\sim20\%$ of present-day passive cluster galaxies with $\logm<10$ must have been ``built into'' the cluster population at very early times.
These findings support a scenario in which clusters mark places where galaxy evolution has been accelerated compared to---but not radically divergent from---the cosmic mean, but are now also in a phase of transforming mainly low-mass systems via environmentally specific phenomena.

We proceed as follows: in Section \ref{sec:data}, we describe the observations and measurements upon which our analysis is based. In Section \ref{sec:result1}, we explore the canonical size--mass relations of our sample and use these to identify important spatiotemporal trends in the data. 
In Section \ref{sec:result2}, we adopt a new framework to reinterpret galaxy structural parameters holistically across all environments, performing a multidimensional analysis similar in spirit to the approach that led to the discovery of the fundamental plane \citep{djorgovski87, dressler87} and the more-fundamental plane \citep{bolton07, auger10} of early-type galaxies. 
We discuss our results in Section \ref{sec:discussion} and summarize in Section \ref{sec:summary}. 
Details of various parts of our analysis are also provided in Appendices. 

Magnitudes are quoted in the AB system \citep{oke83, fukugita96}. We assume $\Omega_m=0.3$, $\Omega_\Lambda=0.7$, $H_0=70\,\kms\, {\rm Mpc}^{-1}$, and a \citet{chabrier03} initial mass function (IMF).
The catalog for galaxy structural parameters are made available as an electronic table associated with this paper and through the GLASS website.\footnote{\url{http://glass.astro.ucla.edu}}


\section{Data}
\label{sec:data}

\subsection{Imaging and Spectroscopy}
\label{sec:basicData}

We base our analysis on HFF imaging and GLASS \hst\ spectroscopy for the first four HFF clusters with complete data: Abell 2744 ($z=0.308$), MACS0416 ($0.396$), MACS0717 ($0.548$), and MACS1149 ($0.544$).
HFF imaging spans ACS F435/606/814W through WFC3IR F105/125/140/160W filters (seven bands), reaching a 5-$\sigma$ limiting point-source depth of $m_{\rm F160W}\approx 28.7$ \citep{kawamata16, lotz16}. 
Both programs use a parallel strategy where WFC3IR and ACS are exposed simultaneously so that the former falls on the cluster core (hereafter ``CLS'') and the latter on a low-density infall/field region (``PR1''). 
HFF provided WFC3IR (ACS) follow-ups on PR1 (CLS), so all photometric data are available in both regions.

GLASS spectroscopy consists of 10-orbit G102 + 4-orbit G141 WFC3 grism observations covering the CLS pointings (containing the vast majority of our cluster sample) from which we derive spectroscopic redshifts (\specz). All PR1 redshifts are photometric (\photz; Section \ref{sec:redshifts}). 
All GLASS spectra are visually inspected for quality. 
Here, we make use only of ``high quality'' redshifts; i.e., those with quality flag $f_{\rm Q}\geq3$ as described in \citet[][]{schmidt14} and \citet{treu15}.

It is worth noting that, due to the limited WFC3IR field of view, our observations probe only the {\it cores} of clusters, i.e., out to $R_{\rm cl}\sim R_{500}\sim0.4~{\rm Mpc}$ at $z\sim0.5$. 
These are, in some sense, the most extreme galaxy environments in the universe, and the locations where environmental processes (e.g., ram pressure stripping) are most effective. 
All GLASS clusters are bright X-ray sources (\citealt{mantz10}), indicating the presence of dense intra-cluster gas.
At the same time, PR1 observations sample close to mean-density environments (``the field''), at least at all redshifts distinct from that of the CLS cluster.
Hence, our sample exhibits almost maximal density contrast, so our analysis should be quite sensitive to environmental effects.

To increase our field sample size, we also include multi-band \hst\ imaging conducted by the {\it eXtreme Deep Field} (XDF) team \citep{illingworth13}.
The XDF encompasses one WFC3IR pointing in GOODS-south, which we use to extend our field galaxy sample.
These data are of comparable depth to the HFF and include the F775W and F850LP filters in addition to the HFF complement.
Ground-based spectroscopic and {\it 3D-HST} grism redshifts \citep{vandokkum13b, momcheva16} are incorporated where available.


\subsection{Photometry and Catalog Construction}
\label{ssec:photz}

After PSF-matching all images to F160W resolution (${\rm FWHM}=0\farcs 18$), we stack the data, weighted by RMS, to maximize detections of faint (low-mass) galaxies. 
This composite image is then run through SExtractor \citep{bertin96} as the detection image.
After removing the intra-cluster light (ICL; see Appendix~\ref{sec:Aa}), photometry is conducted on the individual images based on the detection locations.

To maximize signal-to-noise ratios (S/N) and thus optimize \photz\ estimates, fluxes in each filter are determined within fixed apertures of 12~pixels ($0\farcs7$) in diameter.
When estimating absolute quantities---such as stellar masses---these measurements need to be scaled to account for light outside the aperture.
We do this by adopting {\tt FLUX\_AUTO} from SExtractor as the total flux of each galaxy (see Section~\ref{ssec:sed}).
All photometry is corrected for galactic extinction using the \citet{schlegel98} dust maps.\footnote{\url{https://ned.ipac.caltech.edu/forms/calculator.html}}

Below, we analyze only objects with $m_{\rm F160W}<26\, {\rm mag}$. 
Most of this sample has $S/N_{\rm F160W}>8$, sufficient for accurate stellar mass and structural parameter estimates \citep[][see Appendix \ref{sec:magmass}]{schmidt14b}. 
We have verified that our results are quantitatively robust to this cut-off.
The magnitude criterion corresponds to a mass completeness limit of $\logm\sim7.8$ (Section \ref{ssec:sed}).

\begin{figure}
\begin{center}
	\includegraphics[width=8cm,bb=0 0 288 360]{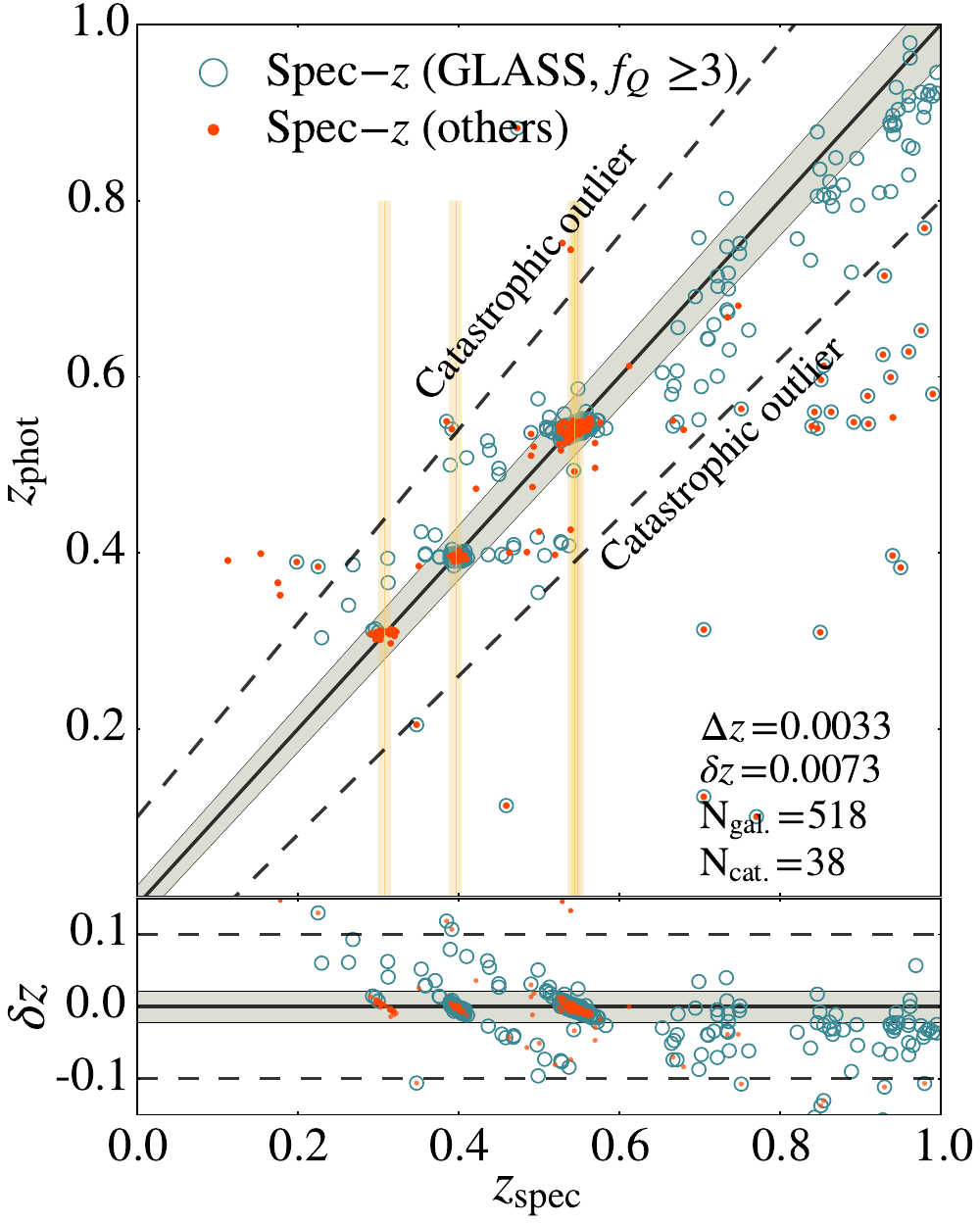}
	\caption{
		Comparison between photometric (\photz) and spectroscopic redshifts (\specz) for 518 objects at $z<1.0$.
		Spectroscopic redshifts are taken from ground-based measurements if available (red points), and from the GLASS catalog otherwise (blue circles; 242 galaxies with quality flag $f_{\rm Q}\geq3$).
		Catastrophic outliers have $| z_\mathrm{spec}-z_\mathrm{phot} |/(1+z_\mathrm{spec}) >0.1$ and lie above or below the dashed lines. 
		These were excluded from the calculation of the median offset, $\Delta z$, and the normalized median absolute deviation, $\delta z$, of the quantity $(z_\mathrm{spec}-z_\mathrm{phot}) /(1+z_\mathrm{spec})$. 
		The gray shaded region corresponds to $3\times\delta z$. 
		Vertical bands mark the redshifts of the four clusters analyzed with widths corresponding to their velocity dispersions ($\sim2000\, \kms$). 
		Our photometric redshifts are in excellent agreement with \specz\ for both cluster and field galaxies.
}
\label{fig:photz}
\end{center}
\end{figure}

\begin{figure}
\begin{center}
	\includegraphics[width=8.0cm,bb=0 0 288 288]{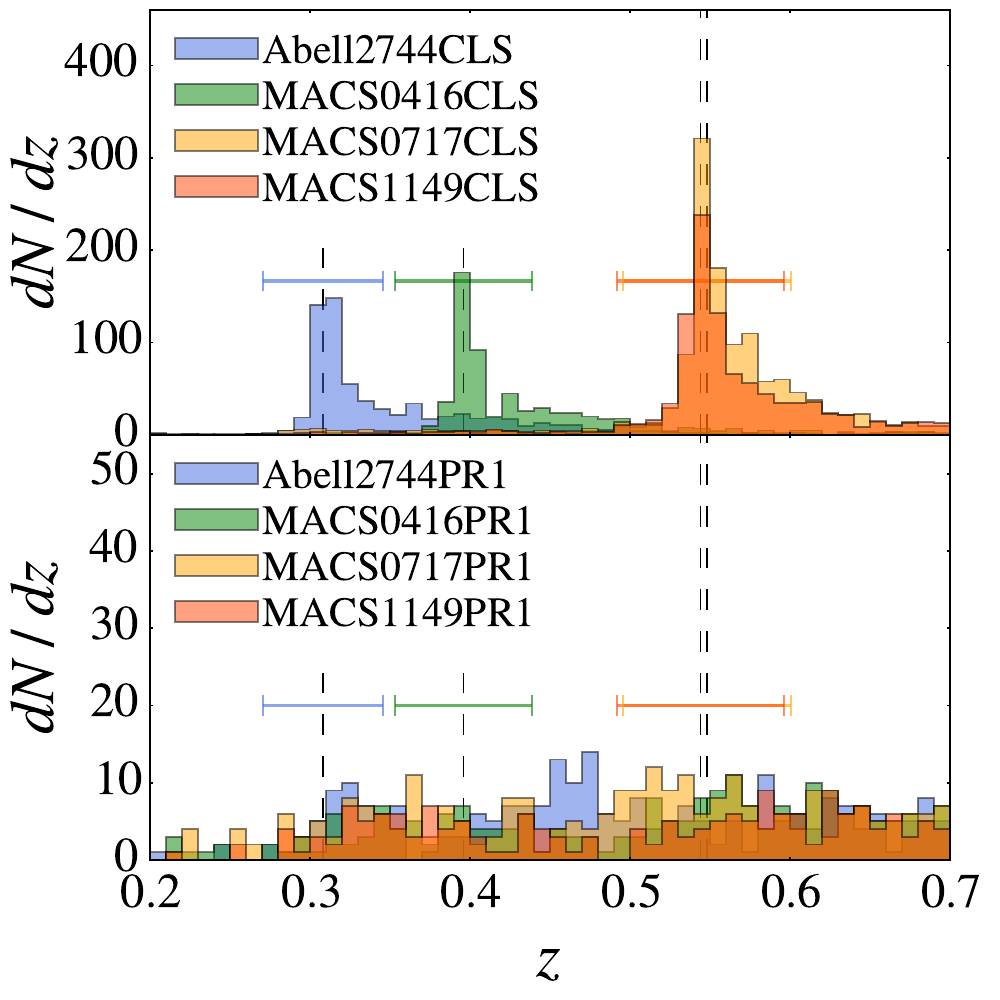}
	\caption{
		Top: \photz\ distributions of galaxies in the four cluster-core (CLS) pointings. 
		Histograms are color coded by cluster, with black dashed lines indicating the spectroscopic mean redshift of each ($z_{\rm cls}$).
		Galaxies within $\pm3\delta z_{\rm phot.}\times(1+z_\mathrm{cls})$ (horizontal bars) are classified as cluster members.
		Bottom: same as the top panel, but for PR1 sources.
		}
\label{fig:photzhist}
\end{center}
\end{figure}


\subsection{Spectroscopic and Photometric Redshifts}
\label{sec:redshifts}

Wherever available, we adopt public spectroscopic redshift (\specz) provided by several authors \citep{owers11, ebeling14, balestra15}.\footnote{\url{http://www.stsci.edu/hst/campaigns/frontier-fields/FF-Data}} 
We supplement these with GLASS \specz.
As mentioned, we include only GLASS redshifts with $f_{\rm Q}\geq3$, corresponding to two or more line detections (e.g., [\ion{O}{3}] and H$\alpha$; \citealt{treu15}). These cover $7\%$ of the $m_{\rm F160W}<26$ sample (269 galaxies). 
GLASS redshifts show excellent agreement with the ground-based measurements.

For galaxies lacking \specz, we derive photometric redshift (\photz) using the seven-band \hst\ photometry discussed above.
We fit all spectral energy distributions (SEDs) using the EAZY code \citep[v.1.01;][]{Brammer08}, implemented with emission line/dusty spectrum templates based on the recipe of \citet{ilbert09}.
Given the depth of the HFF data, $85\%$ of our sample has 3-$\sigma$ detections in more than four bands, supporting reliable \photz\ estimates. 
\subsubsection{Photo-z Priors}
\label{sec:pzPriors}

The only modification we make to the default EAZY fitting routine is to apply different priors for CLS and PR1 objects. 
In PR1, where field galaxies dominate, we apply the default EAZY F160W prior derived from Theoretical
Astrophysical Observatory (TAO) lightcones \citep{bernyk16}.\footnote{\url{https://tao.asvo.org.au/tao/}} 
In CLS, we modify this prior in order to account for the existence of each cluster as follows:
\begin{equation}\label{eq:prior}
	\begin{split}
		p(z\,|\,m_{\rm F160W},\, z_{\rm cls}) = f\times p(z\,|\,m_{\rm F160W})_{\rm fld} \\
			+~(1-f)\times g(z\,|\,z_{\rm cls},\sigma),
	\end{split}
\end{equation}
where $p(z\,|\,m_{\rm F160W})_{\rm fld}$ is the default (field) EAZY prior, $f$ is the fraction of field vs.\ cluster galaxies at a given magnitude (Equation \ref{eq:CLSpriorWeight}), and $g(z_{\rm cls},\sigma)$ is a normalized gaussian centered at the redshift of a given cluster ($z_{\rm cls}$) with a dispersion $\sigma=2000\,\kms$ (about twice the velocity dispersion of each cluster). 
We set $f$ to the number of galaxies in the PR1 pointing over that in the CLS pointing in bins of $m_{\rm F160W}$, assuming that the CLS sample is dominated by cluster members. 
Appendix~\ref{sec:PhotoPrior} provides further details.

We adopt the {\tt z\_peak} EAZY output as our \photz\ estimate.
We compare the best-fit $z_{\rm phot}$ to $z_{\rm spec}$ in Figure \ref{fig:photz}, finding a median offset of $\langle(z_{\rm spec} - z_{\rm phot}) /(1+z_{\rm spec})\rangle = 0.003$, and a median absolute deviation $\delta z_{\rm phot} = 0.0073$ (i.e., 0.7\%) for galaxies at $z<1.0$.
The cluster \photz\ prior is partially responsible for this tight dispersion, but $\delta z_{\rm phot}$ rises to just $1.8\%$ in the PR1 and XDF fields where it is not employed, giving us high confidence in the accuracy of our \photz\ estimates.

Catastrophic outliers are defined to have $|z_\mathrm{spec}-z_\mathrm{phot}|/(1+z_{\rm spec}) >0.1$; they comprise 7.3\% of the \specz\ sample.

After culling to $0.2\leq z \leq0.7$---a redshift range bracketing the four HFF clusters---our final catalog consists of 3948~galaxies, with 2200 in clusters and 1748 in field environments. 
A total of 298 have ground-based and 168 have GLASS spectroscopic redshifts.


\subsection{Cluster and Field Sample Selection}
Cluster members are identified using the redshifts described above. 
As spectroscopic and photometric estimates have different uncertainties, we define different criteria for membership based on the metric:
\begin{equation}
	\delta z_{\rm incl}\equiv\frac{|z - z_{\rm cls}|}{1+z_{\rm cls}}.
\end{equation}
Members have:
\begin{enumerate}
	\item $\delta z_{\rm incl} \leq 0.0084$ or 0.0087 for ground- or space-based \specz, respectively, corresponding to double the typical cluster velocity dispersion (i.e., $\sim2000\, \kms$), convolved with measurement errors. 
	\item $\delta z_{\rm incl} \leq 0.0219$ for \photz, corresponding to $3\times \delta z_{\rm phot}$ (see Section~\ref{sec:redshifts}).
\end{enumerate}

We verified that the selected cluster galaxies mostly belong to the red sequence down to $m_{\rm F160W}\sim25$\,mag in the color--magnitude diagram (not shown), giving us confidence in the selection. 
Figure~\ref{fig:photzhist} shows the sample's redshift distribution.


\subsection{Stellar Masses}
\label{ssec:sed}

Stellar masses for all galaxies are derived from their multi-band photometry (SED fitting), and \specz\ or \photz\ estimate (Section~\ref{sec:redshifts}), using FAST \citep{Kriek09}.
This process requires fluxes to be scaled from aperture measurements ($F_{\rm aper}$; see Section~\ref{ssec:photz}) to the total values ($F_{\rm tot}$) assuming: 
\begin{equation}
F_{\rm tot} = F_{\rm aper} \times {F_{\rm auto}^{\rm F160W}\over F_{\rm aper}^{\rm F160W}},
\end{equation}
where $F_{\rm auto}^{\rm F160W}$ is the total flux ({\tt FLUX\_AUTO}) from SExtractor, covering $3$~Kron radii \citep{kron80}.

This procedure works for $92\%$ of the sample, but for the rest the {\tt FLUX\_AUTO} uncertainties are large enough (mainly due to close, bright neighbors or ICL residuals) that using it risks introducing a bias. 
In these cases---$S/N(F_{\rm auto}^{\rm F160W})\leq1$---we apply no scaling. 
This affects stellar mass estimates very little as the original $0\farcs7$ aperture encompasses $\sim2\,r_{e}$ for most of these systems.

We use the stellar population models of \citet{bruzual03}, assuming solar metallicity, a \citet{chabrier03} IMF, and a \citet{calzetti00} dust law. 
Internal extinction is calculated assuming $0\leq A_V\leq4\, {\rm mag}$ with a grid spacing of $0.1\, {\rm mag}$. 
We adopt exponentially declining star formation histories---${\rm SFR}(t)\propto\exp(-t/\tau)$---with $\log\tau/{\rm yr} \in [8,10]$ in steps of $0.2\, {\rm dex}$. 
Uncertainties are taken as the 1-$\sigma$ limits derived by FAST.


\subsection{Rest-frame Colors}
\label{sec:UVJ}

We wish to examine how environment affects both galaxy structure and star formation. An efficient means of classifying galaxies by their star formation state is by their location in rest-frame $U-V$/$V-J$ (``{\it UVJ}'') color-color space \citep[][]{williams09}. 
These colors are directly calculated by convolving the best fit EAZY spectral templates with Johnson $U,\, V$, and 2MASS $J$-band filters. 
We follow \citet{williams09} and define red (quiescent) galaxies to have:
\begin{equation}
	\begin{aligned}
		U-V >0.88 &\times(V-J)+0.69,\\
	 	U-V &>1.3,\\
	 	V-J &<1.6.
	\end{aligned}
\end{equation}
We refer to all others as blue (star-forming) galaxies.

\begin{figure}
\begin{center}
	\includegraphics[width=0.49\textwidth]{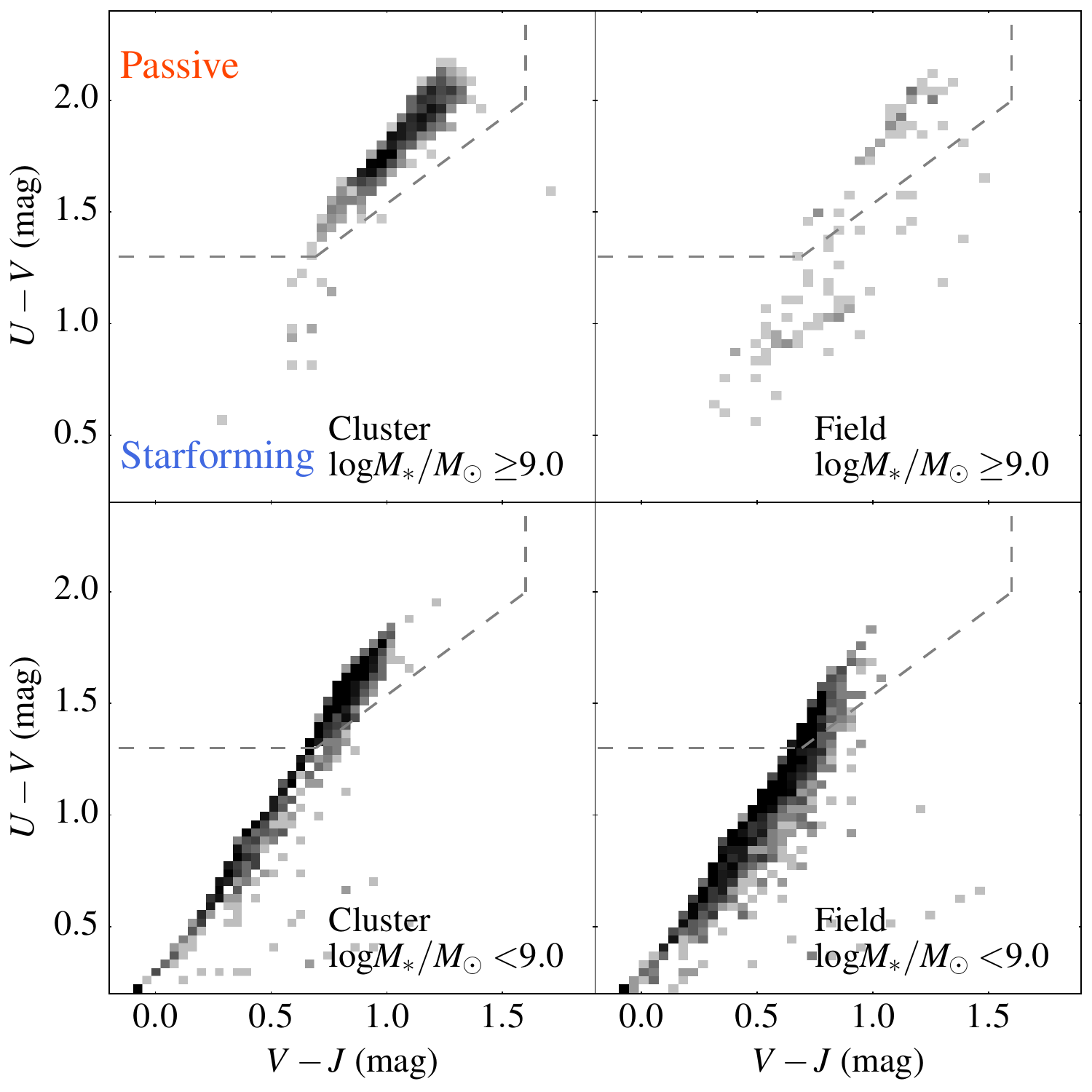}
	\caption{
		Rest-frame {\it UVJ} diagrams for cluster (left) and field (right) galaxies, with masses $\logm\geq9$ (top) and $<9$ (bottom).
		Galaxies in the top-left box (dashed lines; from \citealt{williams09}) in each panel are classified as passive; the remaining galaxies we define as star-forming.
		A clear trend of increasing passive fraction with both stellar mass and environmental density emerges as expected, but we also find low-mass passive cluster galaxies to be redder than those in the field. 
		}
\label{fig:uvj}
\end{center}
\end{figure}

Figure~\ref{fig:uvj} shows the distribution of our sample---split by mass ($\logm\lessgtr9$) and environment---in {\it UVJ} space. 
We see not only the bimodality of passive/star-forming galaxies, but also the trend of increasing passive fraction with stellar mass, as expected. 
We note also that, at low stellar mass, cluster passive galaxies appear slightly redder than their field counterparts. 
We return to this point in Section \ref{ssec:slmass}.

\begin{figure}
\begin{center}
	\includegraphics[width=0.49\textwidth, trim = 0cm 1.4cm 0cm 0cm]{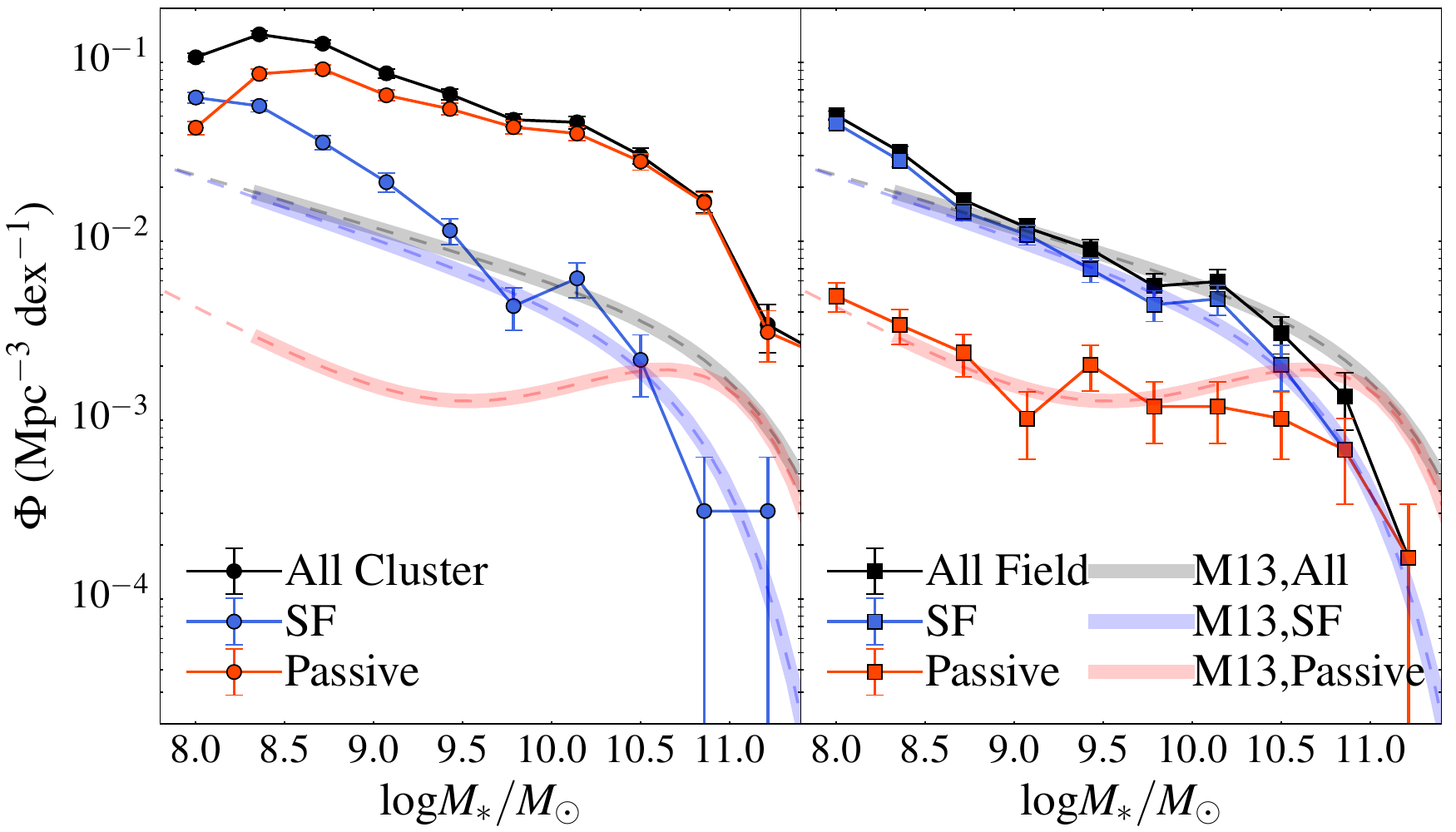}
	\includegraphics[width=0.49\textwidth, trim = 0cm 0cm 0cm 0.05cm]{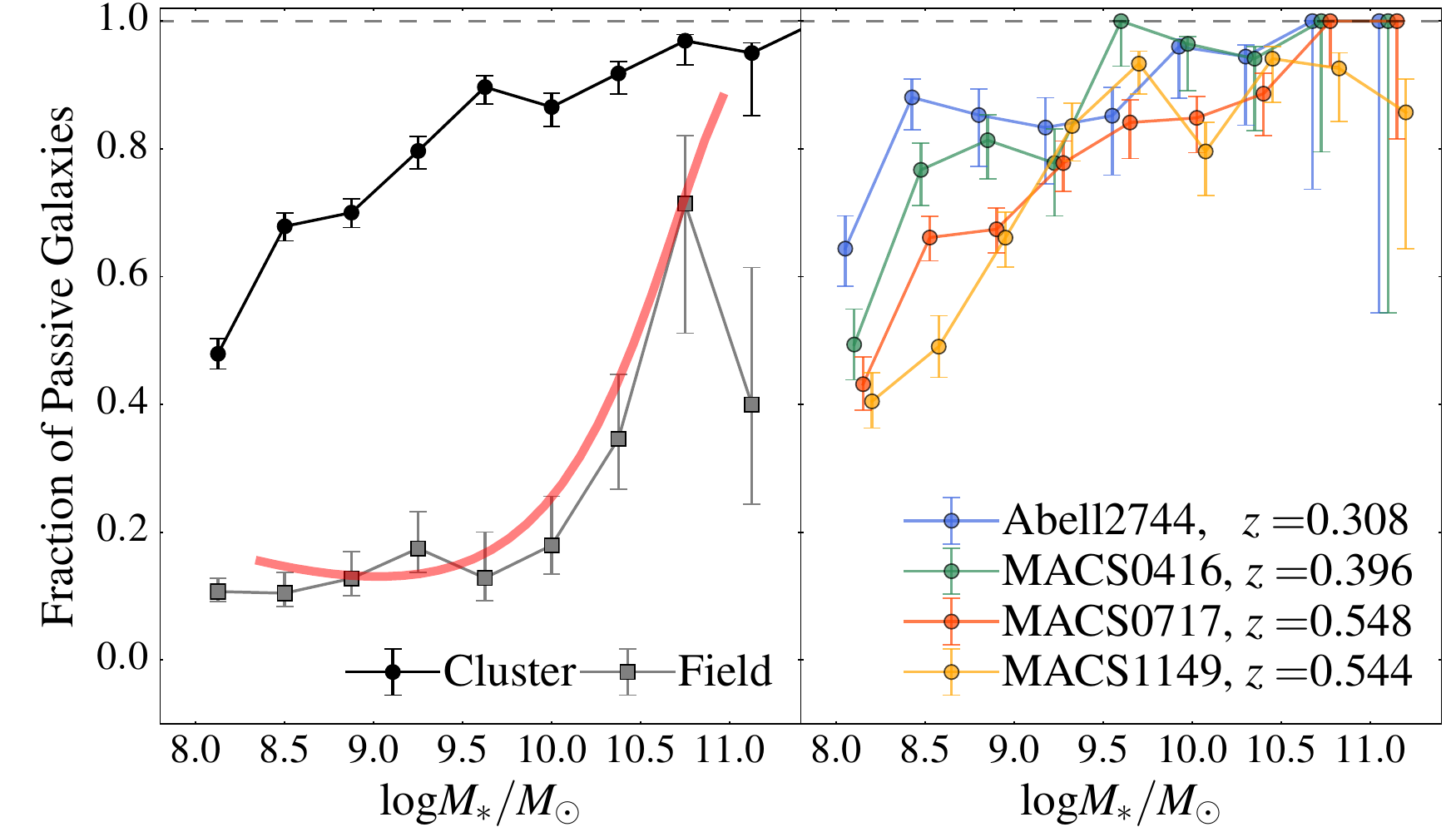}
	\caption{
		Top: stellar mass functions of cluster members (left) and field galaxies (right), color-coded by population (black = all, blue = star-forming (SF), red = passive).
		Error bars assume a binomial distribution.
		Results from \citet{muzzin13} for field galaxies at $0.2<z<0.5$ are shown for comparison (colored consistently). 
		Bottom left: fraction of passive galaxies as a function of stellar mass for cluster members (black) and field galaxies (gray).
		The \citet{muzzin13} field result is shown as the red solid line.
		Our measurement diverges from those authors' at the highest stellar mass, but our sample of such objects is very small.
		Bottom right: fraction of passive galaxies for each of our four clusters.
		The population of low-mass galaxies ($\logm<9$) evolves even in the narrow redshift range these span (or $\sim2$ Gyr of cosmic time).
		}
\label{fig:smf}
\end{center}
\end{figure}


\subsection{Cross-checking Photo-z Accuracy: Galaxy Number Counts and Passive/Star-forming Fractions}
\label{ssec:smf}

Figure \ref{fig:smf}, top, shows the stellar mass distributions of our samples. 
Our purpose here is not to investigate, e.g., the best-fit Schechter parameters, but rather to demonstrate the robustness of our largely \photz\ based cluster/field sample separation.

Comparing our field results (right) with those from \citet{muzzin13} in a similar redshift range ($0.2<z<0.5$), we see good consistency at $\logm\geq8.3$, the completeness limit of the previous study.
This is true for all galaxies, and both passive and star-forming sub-classes, suggesting that our field sample is indeed representative of the general galaxy population and not excessively contaminated by cluster objects.

Figure \ref{fig:smf}, bottom-left, shows the fraction of passive galaxies for cluster and field environments as a function of stellar mass. There is a significant excess in clusters over the entire mass range, rising from $\sim50\%$ at $\logm\approx8$ to over $90\%$ at $\logm>10$. This is also as expected from previous spectroscopic studies \citep[e.g.,][their Figure 16]{dressler13}, and suggests that our cluster sample is not excessively diluted by field galaxies. 

Interestingly, the bottom-right panel in Figure \ref{fig:smf} shows that, even within our sample's rather narrow redshift range ($0.3 \lesssim z_{\rm cls} \lesssim 0.6$), evolution in the cluster passive fraction---the \cite{butcher78} Effect---is detected.
The evolution is observed only for low-mass systems ($\logm<9.0$), which might suggest that environmental effects are most pronounced for these systems.
We combine this result with our inferred galaxy structural properties to develop a preferred scenario for the origin and evolution of these low-mass passive cluster systems in Section \ref{sec:discussion}, but regardless: all of the above results suggest that our field/cluster galaxy sample selection process is accurate.

\tabletypesize{\footnotesize} \tabcolsep=0.1cm
\begin{deluxetable*}{lccccccccccccccc} \tablecolumns{21}
    \tablewidth{0pt} 
    \tablecaption{GLASS Size--Mass Relations: Source and Structural Catalog}
    \tablehead{\colhead{ID$_{\rm cls}$} & \colhead{ID$_{\rm id}$} & \colhead{$\alpha_\textrm{J2000}$} &
    \colhead{$\delta_\textrm{J2000}$} & \colhead{$z_{\rm best}$} &
    \colhead{F160W mag} & \colhead{$\logm$} & \colhead{$\log r_e/\,{\rm kpc}$} & 
    \colhead{$\log n$} & \colhead{$\log b/a$} & 
    \colhead{$f_{\tt GALFIT}$} & \colhead{$(U-V)\,{\rm mag}$} & \colhead{$\Sigma_{\rm 5th}\,{\rm Mpc^{-2}}$}
    }
\startdata
\hline
1 & 1227 & 3.57507563 & -30.37707701 & 0.31 & 18.43 $\pm$ 0.10 & 10.44 $\pm$ 0.00 & 0.69 $\pm$ 0.05 & 0.56 $\pm$ 0.05 & -0.48 $\pm$ 0.05 & 0 & 2.00 & 414.47\\
1 & 1228 & 3.57344998 & -30.37793406 & 0.31 & 19.84 $\pm$ 0.10 & 9.85 $\pm$ 0.00 & 0.39 $\pm$ 0.00 & 0.66 $\pm$ 0.00 & -0.27 $\pm$ 0.05 & 0 & 1.89 & 614.00\\
1 & 1231 & 3.57513621 & -30.37852385 & 0.31 & 21.42 $\pm$ 0.10 & 9.12 $\pm$ 0.00 & 0.61 $\pm$ 0.01 & 0.61 $\pm$ 0.02 & -0.10 $\pm$ 0.01 & 0 & 1.72 & 969.51\\
1 & 1694 & 3.57754208 & -30.37887163 & 0.31 & 21.15 $\pm$ 0.10 & 9.31 $\pm$ 0.00 & 0.14 $\pm$ 0.00 & 0.19 $\pm$ 0.00 & -0.19 $\pm$ 0.05 & 0 & 1.83 & 652.59\\
1 & 1711 & 3.60954388 & -30.38211056 & 0.29 & 17.93 $\pm$ 0.10 & 10.57 $\pm$ 0.00 & 0.22 $\pm$ 0.00 & 0.43 $\pm$ 0.00 & -0.14 $\pm$ 0.05 & 0 & 2.07 & 10.54\\
1 & 1720 & 3.58980645 & -30.37841131 & 0.33 & 24.62 $\pm$ 0.17 & 7.90 $\pm$ 0.03 & 0.21 $\pm$ 0.00 & 0.32 $\pm$ 0.01 & -0.15 $\pm$ 0.01 & 0 & 1.47 & 36.94\\
1 & 1756 & 3.58918156 & -30.37894909 & 0.31 & 22.77 $\pm$ 0.11 & 8.58 $\pm$ 0.00 & 0.18 $\pm$ 0.00 & 0.23 $\pm$ 0.00 & -0.09 $\pm$ 0.05 & 0 & 1.49 & 129.73\\
1 & 1763 & 3.59763710 & -30.37920224 & 0.31 & 20.41 $\pm$ 0.10 & 9.60 $\pm$ 0.00 & 0.32 $\pm$ 0.00 & 0.49 $\pm$ 0.00 & -0.38 $\pm$ 0.05 & 0 & 1.82 & 110.44\\
1 & 1797 & 3.57885731 & -30.37943037 & 0.31 & 23.85 $\pm$ 0.12 & 8.15 $\pm$ 0.01 & 0.01 $\pm$ 0.00 & 0.04 $\pm$ 0.01 & -0.01 $\pm$ 0.00 & 0 & 1.53 & 374.08\\
1 & 1804 & 3.57972596 & -30.37953693 & 0.31 & 23.32 $\pm$ 0.12 & 8.38 $\pm$ 0.01 & 0.20 $\pm$ 0.00 & 0.08 $\pm$ 0.00 & -0.26 $\pm$ 0.05 & 0 & 1.52 & 238.31\\
1 & 1823 & 3.57167798 & -30.37970624 & 0.31 & 23.53 $\pm$ 0.11 & 8.30 $\pm$ 0.01 & -0.27 $\pm$ 0.00 & 0.25 $\pm$ 0.01 & -0.05 $\pm$ 0.00 & 0 & 1.68 & 614.00\\
1 & 1830 & 3.59539540 & -30.38040287 & 0.31 & 19.45 $\pm$ 0.10 & 9.92 $\pm$ 0.00 & 0.54 $\pm$ 0.00 & 0.49 $\pm$ 0.00 & -0.27 $\pm$ 0.05 & 0 & 1.72 & 271.21\\
1 & 1853 & 3.57922360 & -30.38018704 & 0.30 & 23.61 $\pm$ 0.12 & 8.21 $\pm$ 0.01 & 0.21 $\pm$ 0.00 & -0.01 $\pm$ 0.01 & -0.14 $\pm$ 0.01 & 0 & 1.57 & 44.58\\
1 & 1879 & 3.57579814 & -30.38040518 & 0.31 & 23.70 $\pm$ 0.12 & 8.20 $\pm$ 0.01 & 0.10 $\pm$ 0.00 & 0.14 $\pm$ 0.01 & -0.12 $\pm$ 0.01 & 0 & 1.50 & 667.78\\
1 & 1920 & 3.57119702 & -30.38093715 & 0.31 & 23.26 $\pm$ 0.12 & 8.42 $\pm$ 0.01 & 0.13 $\pm$ 0.00 & 0.19 $\pm$ 0.01 & -0.16 $\pm$ 0.01 & 0 & 1.64 & 294.77\\
1 & 1933 & 3.57785827 & -30.38120896 & 0.34 & 23.61 $\pm$ 0.12 & 8.13 $\pm$ 0.01 & 0.11 $\pm$ 0.00 & 0.22 $\pm$ 0.01 & -0.48 $\pm$ 0.05 & 0 & 1.34 & 20.75\\
1 & 1934 & 3.57847019 & -30.38131772 & 0.31 & 20.97 $\pm$ 0.10 & 9.38 $\pm$ 0.00 & -0.09 $\pm$ 0.05 & 0.45 $\pm$ 0.00 & -0.19 $\pm$ 0.05 & 0 & 1.82 & 483.69\\
1 & 1951 & 3.57942272 & -30.38139612 & 0.32 & 23.83 $\pm$ 0.13 & 8.20 $\pm$ 0.01 & 0.07 $\pm$ 0.00 & 0.27 $\pm$ 0.01 & -0.02 $\pm$ 0.00 & 0 & 1.49 & 136.01\\
1 & 1980 & 3.57086899 & -30.38200261 & 0.31 & 20.30 $\pm$ 0.10 & 9.57 $\pm$ 0.00 & 0.15 $\pm$ 0.00 & 0.32 $\pm$ 0.05 & -0.46 $\pm$ 0.05 & 0 & 1.72 & 387.68\\
1 & 1984 & 3.59027996 & -30.38269453 & 0.30 & 19.79 $\pm$ 0.10 & 9.80 $\pm$ 0.00 & 0.26 $\pm$ 0.00 & 0.53 $\pm$ 0.00 & -0.36 $\pm$ 0.05 & 0 & 1.99 & 40.39\\
1 & 2006 & 3.58803926 & -30.38255939 & 0.31 & 20.11 $\pm$ 0.10 & 9.67 $\pm$ 0.00 & 0.42 $\pm$ 0.00 & 0.32 $\pm$ 0.05 & -0.36 $\pm$ 0.05 & 0 & 1.87 & 167.35\\
1 & 2029 & 3.60688322 & -30.38226934 & 0.32 & 24.65 $\pm$ 0.14 & 7.86 $\pm$ 0.02 & -0.08 $\pm$ 0.00 & -0.04 $\pm$ 0.05 & -0.30 $\pm$ 0.02 & 0 & 1.45 & 57.93\\
1 & 2031 & 3.56534751 & -30.38294889 & 0.30 & 19.29 $\pm$ 0.10 & 10.16 $\pm$ 0.00 & 0.03 $\pm$ 0.00 & 0.53 $\pm$ 0.00 & -0.28 $\pm$ 0.05 & 0 & 2.01 & 47.15\\
1 & 2056 & 3.59302001 & -30.38296723 & 0.31 & 22.45 $\pm$ 0.11 & 8.73 $\pm$ 0.00 & 0.22 $\pm$ 0.01 & 0.18 $\pm$ 0.03 & -0.13 $\pm$ 0.02 & 0 & 1.53 & 281.51\\
1 & 2062 & 3.57582476 & -30.38339820 & 0.30 & 23.33 $\pm$ 0.12 & 8.35 $\pm$ 0.01 & 0.03 $\pm$ 0.01 & -0.01 $\pm$ 0.11 & -0.02 $\pm$ 0.05 & 0 & 1.62 & 124.27\\
1 & 2063 & 3.57439610 & -30.38365253 & 0.30 & 18.14 $\pm$ 0.10 & 10.48 $\pm$ 0.00 & 0.60 $\pm$ 0.00 & 0.43 $\pm$ 0.00 & -0.07 $\pm$ 0.05 & 0 & 2.00 & 147.80\\
1 & 2073 & 3.57083471 & -30.38293109 & 0.31 & 23.64 $\pm$ 0.12 & 8.23 $\pm$ 0.01 & -0.03 $\pm$ 0.00 & -0.04 $\pm$ 0.02 & -0.15 $\pm$ 0.01 & 0 & 1.54 & 248.03\\
1 & 2088 & 3.58371863 & -30.38466949 & 0.31 & 19.61 $\pm$ 0.10 & 9.83 $\pm$ 0.00 & 0.50 $\pm$ 0.00 & 0.50 $\pm$ 0.05 & -0.10 $\pm$ 0.05 & 0 & 1.63 & 213.00\\
1 & 2133 & 3.56869016 & -30.38354809 & 0.32 & 24.00 $\pm$ 0.42 & 8.19 $\pm$ 0.13 & -0.23 $\pm$ 0.02 & -0.17 $\pm$ 0.29 & -0.09 $\pm$ 0.07 & 0 & 1.58 & 59.01\\
1 & 2161 & 3.59254902 & -30.38531028 & 0.31 & 19.54 $\pm$ 0.10 & 9.94 $\pm$ 0.00 & -0.02 $\pm$ 0.05 & 0.54 $\pm$ 0.00 & -0.34 $\pm$ 0.05 & 0 & 2.00 & 311.00\\
1 & 2171 & 3.57104894 & -30.38815432 & 0.31 & 21.45 $\pm$ 0.11 & 9.18 $\pm$ 0.00 & 0.27 $\pm$ 0.00 & 0.26 $\pm$ 0.00 & -0.04 $\pm$ 0.05 & 0 & 1.74 & 294.73\\
-- & -- & -- & -- & -- &  -- & --  & -- & -- & -- & -- & -- & --\\
99 & 1188 & 53.14167849 & -27.77312189 & 0.51 & 23.81 $\pm$ 0.17 & 8.49 $\pm$ 0.03 & 0.42 $\pm$ 0.00 & 0.16 $\pm$ 0.01 & -0.14 $\pm$ 0.01 & 0 & 0.76 & 1.05\\
99 & 1198 & 53.15121278 & -27.77283955 & 0.63 & 23.67 $\pm$ 0.11 & 8.47 $\pm$ 0.00 & 0.23 $\pm$ 0.00 & 0.13 $\pm$ 0.01 & -0.46 $\pm$ 0.05 & 0 & 0.65 & 2.56\\
99 & 1247 & 53.13980735 & -27.77327735 & 0.28 & 21.06 $\pm$ 0.10 & 9.44 $\pm$ 0.00 & 0.36 $\pm$ 0.00 & -0.43 $\pm$ 0.01 & -0.36 $\pm$ 0.05 & 0 & 1.13 & 2.21\\
99 & 1249 & 53.16857816 & -27.77293191 & 0.59 & 25.68 $\pm$ 0.13 & 7.95 $\pm$ 0.01 & -0.29 $\pm$ 0.01 & 0.60 $\pm$ 0.08 & -0.22 $\pm$ 0.04 & 0 & 0.81 & 1.81\\
99 & 1252 & 53.17501343 & -27.77289971 & 0.67 & 26.15 $\pm$ 0.16 & 7.86 $\pm$ 0.02 & -0.02 $\pm$ 0.02 & 0.21 $\pm$ 0.24 & -0.43 $\pm$ 0.13 & 0 & 1.04 & 0.81\\
99 & 1300 & 53.16196274 & -27.77391726 & 0.29 & 22.39 $\pm$ 0.10 & 8.48 $\pm$ 0.00 & 0.34 $\pm$ 0.00 & 0.07 $\pm$ 0.01 & -0.25 $\pm$ 0.01 & 0 & 0.71 & 6.32\\
99 & 1302 & 53.16080932 & -27.77538208 & 0.62 & 20.86 $\pm$ 0.10 & 10.14 $\pm$ 0.00 & 0.74 $\pm$ 0.00 & 0.02 $\pm$ 0.05 & -0.33 $\pm$ 0.05 & 0 & 1.35 & 2.60\\
99 & 1303 & 53.16016086 & -27.77552891 & 0.62 & 20.87 $\pm$ 0.10 & 10.18 $\pm$ 0.00 & 0.26 $\pm$ 0.00 & 0.75 $\pm$ 0.00 & -0.06 $\pm$ 0.05 & 0 & 1.86 & 2.60\\
99 & 1304 & 53.16233363 & -27.77505586 & 0.42 & 20.25 $\pm$ 0.10 & 10.06 $\pm$ 0.00 & 0.47 $\pm$ 0.00 & 0.05 $\pm$ 0.05 & -0.07 $\pm$ 0.05 & 0 & 0.97 & 1.30\\
99 & 1410 & 53.15074871 & -27.77433715 & 0.58 & 23.14 $\pm$ 0.11 & 8.51 $\pm$ 0.00 & 0.45 $\pm$ 0.00 & -0.24 $\pm$ 0.02 & -0.28 $\pm$ 0.01 & 0 & 1.07 & 1.84\\

\enddata   
\tablecomments{
(a): Cluster ID. 1: Abell2744CLS 2: MACS0416CLS 3: MACS0717CLS 4: MACS1149CLS 5: Abell2744PR1 6: MACS0416PR1 7: MACS0717PR1 8: MACS1149PR1 99: XDF.
(b): ID for individual objects.
(c): $z_best$ is ground-based spectroscopic redshift if available, implemented with GLASS grism redshift, and photometric-redshift derived by EAZY for else.
(d): F160W-band Auto magnitude derived by SExtractor.
(e): Stellar mass derived by FAST.
(f): F160W-band structural parameters derived by GALFIT.
(g): Visual inspection flag for sources $2\sigma$ above/below the size-mass relation for each population. 0: Fine 1: Contaminated 2:Point sources.
(h): Rest-frame $U-V, V-J$ colors derived by convolving best-fit templates by EAZY.
(i): Fifth-closest local galaxy number density.
}
\label{tab3}
\end{deluxetable*}


\begin{figure*}
\begin{center}
	\includegraphics[width=0.9\textwidth]{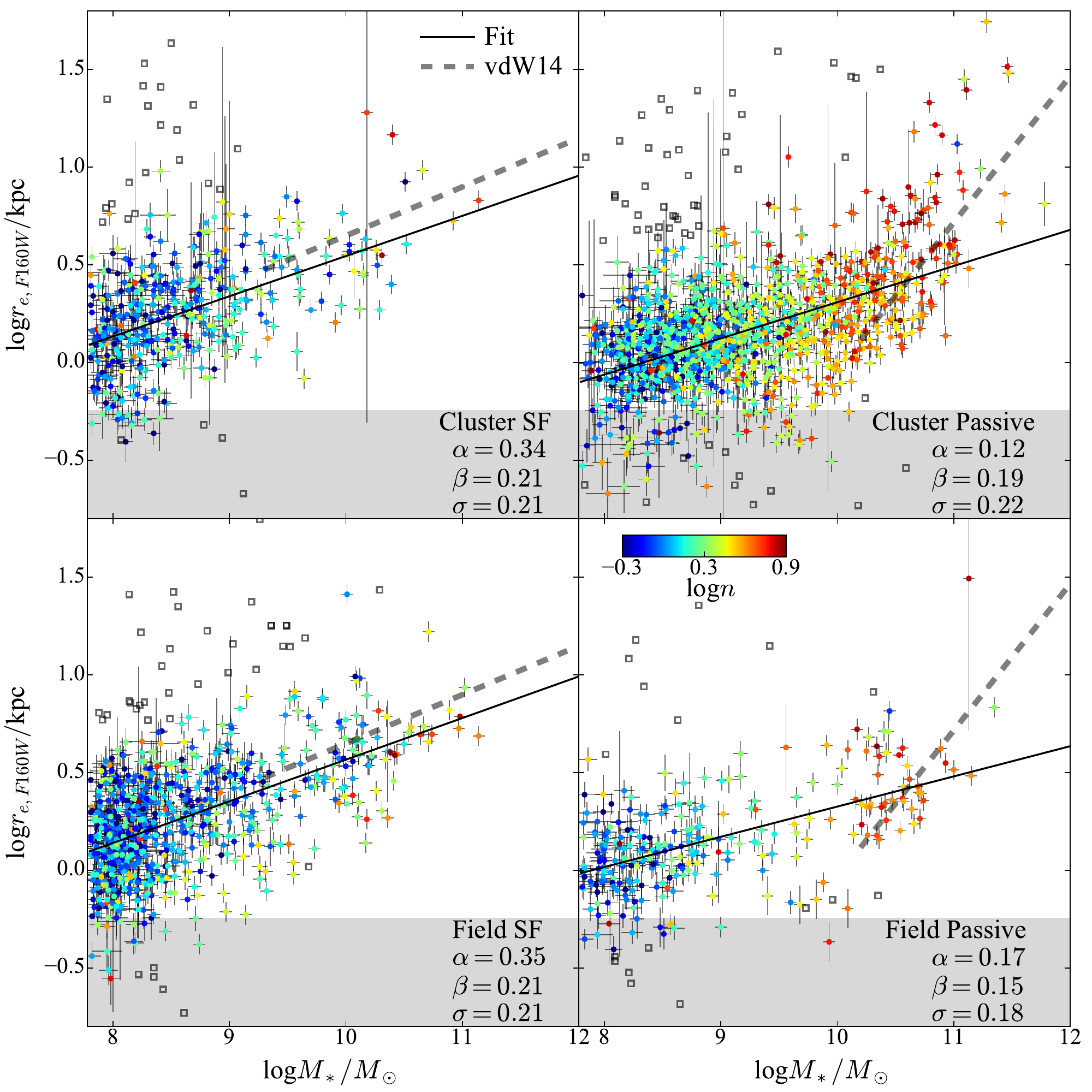}
	\caption{
		The F160W size--mass relation of SF and passive galaxies (left/right, \resp) in clusters and the field (top/bottom).
		Colors reflect galaxy F160W S\'ersic indices ($\log n$).
		Open squares denote systems with close, large/bright neighbors, which comprise 132 visually inspected outliers, and are excluded from further analysis (Section~\ref{sec:parameters}).
		 As Figure \ref{fig:2slopes} shows, low- and high-mass passive galaxies have markedly different size--mass relations, hence, due to our inclusion of galaxies with masses as low as $\logm=7.8$, the slopes we obtain for passive galaxies are $\sim0.3\, {\rm dex}$ per dex shallower than previous estimates \citep[e.g.,][overplotted as black dashes over their fitting range]{vanderwel14}.  
		Size estimates are robust above gray zones at the bottom of all plots, showing $r_{e}\leq{\rm FWHM_{F160W}}/2$ at $z = 0.5$ \citep{morishita14}.
		}
\label{fig:MSR}
\end{center}
\end{figure*}

\subsection{Structural Parameters}
\label{sec:parameters}

Galaxy structural parameters---half-light radii ($r_{e}$), axis ratios ($q\equiv$ semi-minor$/$semi-major axis), and S\'ersic indices ($n$)---are estimated by fitting single S\'ersic profiles \citep{sersic63} using GALFIT \citep{pengGALFIT}. 
Initial guesses for the relevant parameters derive from the SExtractor output. 

Although we discuss only the F160W structural parameters here---corresponding to rest-frame wavelengths of $\sim1.0~\mu$m at $z\sim0.5$---we perform fits in all {\it HST} bands to consistently estimate the ICL properties (Appendix~\ref{sec:Aa}).
After subtracting the ICL from the original CLS image, we then re-estimate the structural parameters for those galaxies and adopt the second-round values. In PR1, we adopt the initial fitting results.

During fitting, we constrain centroids and magnitudes to within $3\,{\rm pixels}$ (in $x$ and $y$) and 1 mag of the SExtractor input values. 
We also set $1<r_e/{\rm pixel} <150$ ($0.4<r_e/{\rm kpc}<60$ at $z\sim0.5$), $0.1<n<8$ (S\'ersic index), and $q>0.2$.
``Successful'' fits are those whose derived parameters fall within these limits. 
Failures are excluded from further analysis.
Close-neighbors---objects with centroids within $6\arcsec$ of target galaxies---are fit simultaneously.

We also visually inspect outliers which reside 2-$\sigma$ above/below the size--mass relations of each population (see next section), and exclude 132 galaxies whose fits are substantially affected by proximity to very bright galaxies/belong to a blended pair, or have grossly distorted morphologies.
Our final catalog contains 2636 galaxies with robust structural parameters. 
This corresponds to a mean success rate of $\sim68\%$, rising from $\sim64\%$ at $\logm\sim8.0$ to $>80\%$ at $\logm\sim10.0$. 
Appendix~\ref{sec:magmass} provides further details of the fitting procedure.
Structural properties, with SED fitting parameters, are shown in Table~\ref{tab3} and available online.


\begin{figure}
\begin{center}
	\includegraphics[width=0.49\textwidth]{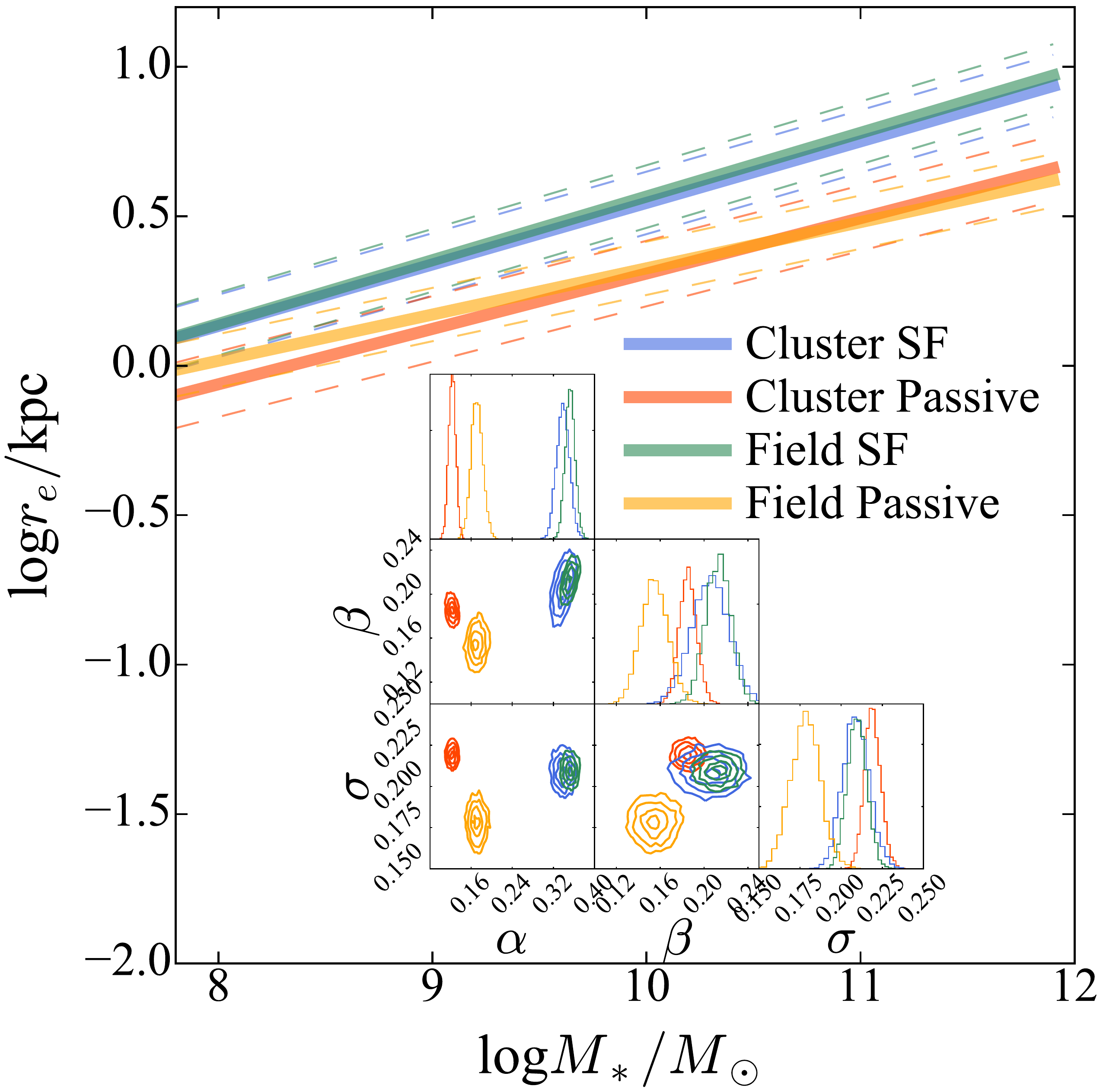}
	\caption{
		Comparison of the best-fit size--mass relations (Equation \ref{eq:canonicalSMR}) for the four populations considered in Figure \ref{fig:uvj}: cluster passive and star-forming galaxies (red/blue lines); field passive and star-forming galaxies (orange/green lines). 
		Dashed lines show the inferred intrinsic dispersion of the relations, $\sigma$.
		Comparison of the best-fit parameters is shown in the inset.
		Contours reflect $68\%$, $96\%$, and $99\%$ confidence intervals as determined using an MCMC solver.
		}
\label{fig:slope}
\end{center}
\end{figure}

\begin{figure}
\begin{center}
\includegraphics[width=0.49\textwidth]{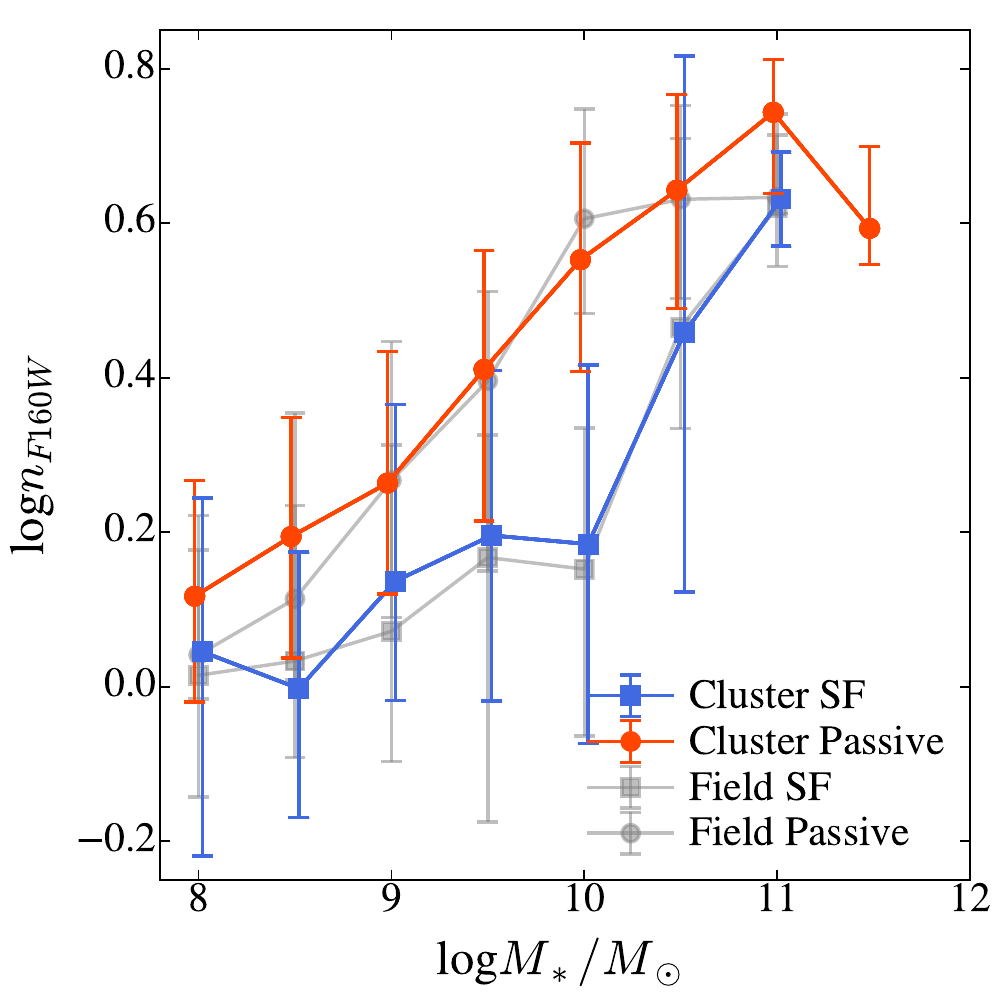}
\caption{
Median and 16th--84th percentile spreads of F160W S\'ersic index, $n$, for cluster star-forming/passive galaxies in bins of stellar mass.
Background gray plots are those for field star-forming/passive galaxies.
Values for cluster galaxies are replotted in light gray in the right panel for comparison.
Star-forming galaxies have similar $n$ distributions at $\logm\lesssim10$, while passive galaxies display a monotonically rising trend.
}
\label{fig:sersic}
\end{center}
\end{figure}

\begin{figure}
\begin{center}
	\includegraphics[width=0.49\textwidth]{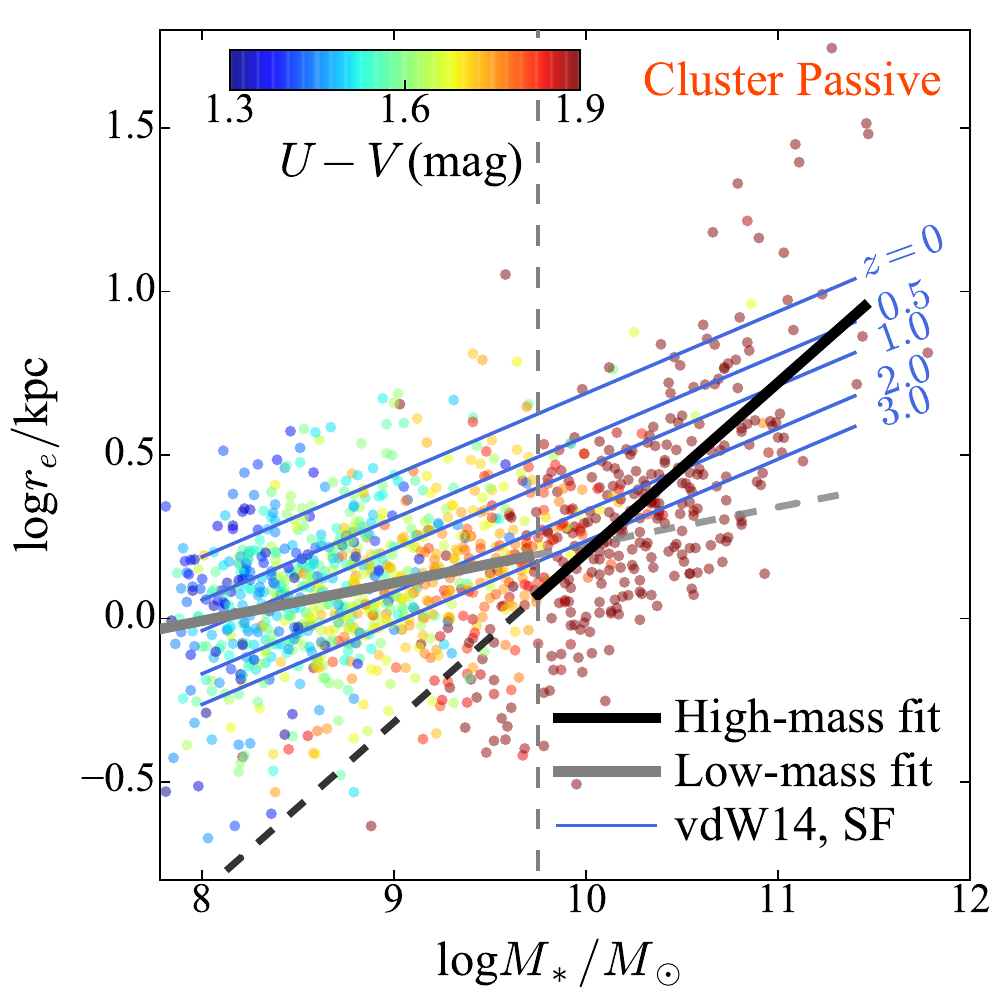}
	\caption{
		Size--mass relation for passive cluster galaxies, same as the right-top panel in Figure~\ref{fig:MSR}, but points are color-coded by rest-frame $U-V$ color.
		We separately fit low- and high-mass systems---split at $\logm=9.8$ (vertical dashed line), where S\'ersic index of star-forming galaxies deviate (Figure~\ref{fig:sersic}).
		The low-mass slope (gray line) is almost flat ($\beta_1=0.11$), while the high-mass fit (black line) is much steeper ($\beta_2=0.51$).
		 Size--mass slopes of {\it star-forming galaxies} by \citet{vanderwel14} are overlaid to discuss the star-forming population as a parent sample of passive galaxies.
		}
\label{fig:2slopes}
\end{center}
\end{figure}


\begin{figure*}
\begin{center}
	\includegraphics[width=0.49\linewidth]{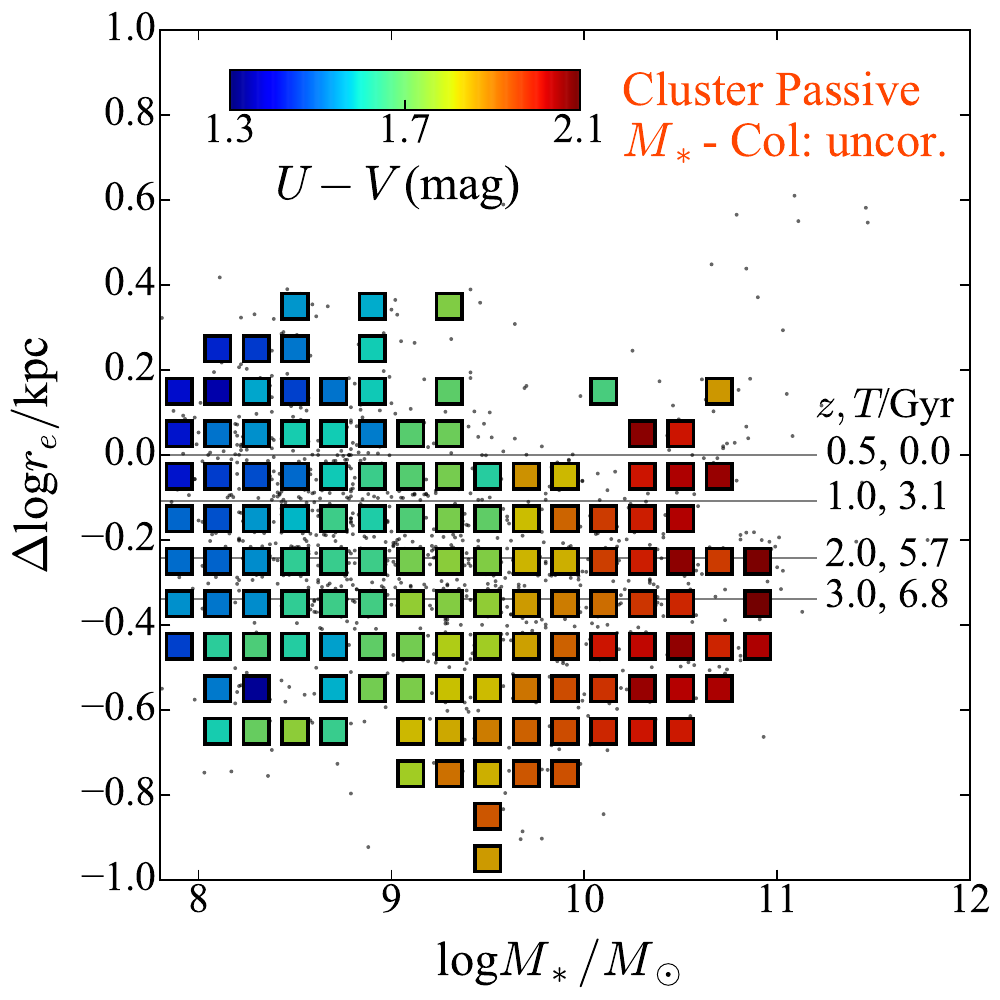}
	\includegraphics[width=0.49\linewidth]{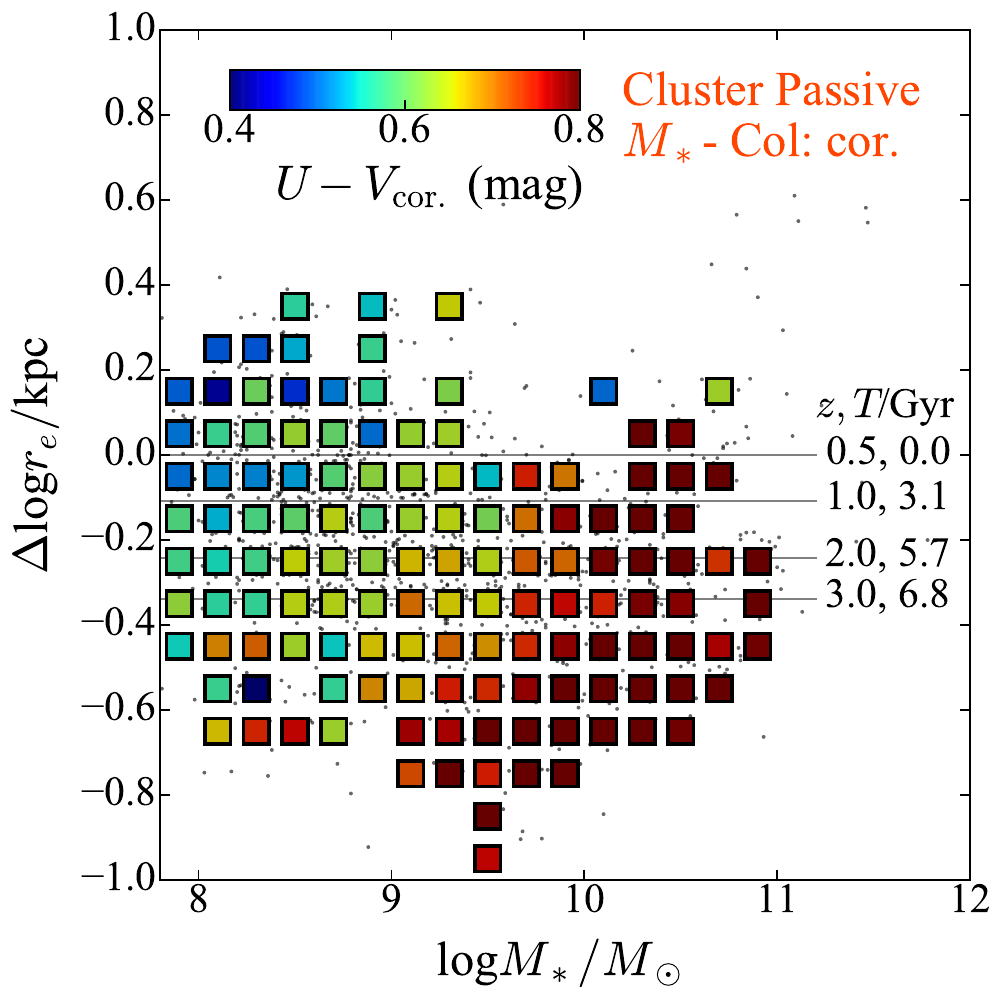}
	\caption{
		Left:
		size differences, $\Delta \log r_e$, of passive cluster galaxies from the fit for SF galaxies (points).
		The horizontal lines are sizes of SF galaxies at different redshifts (printed with $T$~Gyr from the epoch of observation, $z\sim0.45$; same as blue lines in Figure~\ref{fig:2slopes}) from \citet{vanderwel14}.
		Median values for each sub-space are shown by the filled squares, color-coded by $U-V$ color, which serves as an age indicator.
		These show redder colors to correspond with larger offsets from the mean size of SF galaxies at $z\sim0.5$, suggesting both that (1) smaller-size galaxies are older, and (2) only the {\it largest}-size low-mass passive galaxies are consistent with having recently been drawn from the SF population.
		Right:
		same as the left but $U-V$ color is corrected for the intrinsic color--mass relation of SF (putative progenitor) galaxies with Eq.~\ref{EqUVcor}.
		The color trends along the offset persist after the correction.
}
\label{fig:lowQG}
\end{center}
\end{figure*}

\begin{figure}
\begin{center}
	\includegraphics[width=0.98\linewidth]{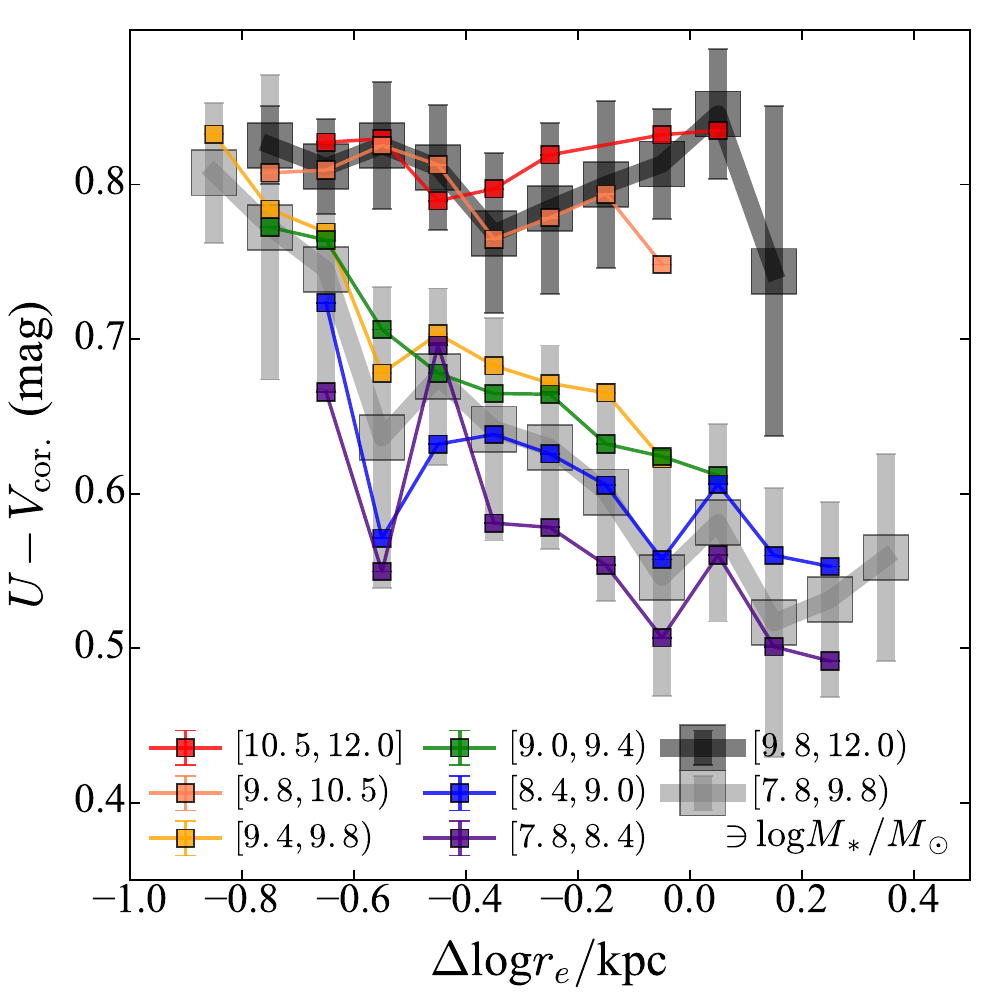}
	\caption{Color trend ($U-V_{\rm cor.}$) along $\Delta \log r_e$ within given stellar mass bins (colored squares), observed in Figure~\ref{fig:lowQG}.
	The medians for high-mass ($\logm\geq9.8$) and low-mass ($\logm<9.8$) galaxies are shown with larger symbols (black and gray, respectively).
	The error bars are median absolute deviations.
	We see a bimodal trend at larger $\Delta \log r_e$ values, where low-mass galaxies are recently quenched from star-forming population, while at smaller $\Delta \log r_e$ values the bimodality is converged.
}
\label{fig:lowQG2}
\end{center}
\end{figure}

\tabletypesize{\footnotesize} \tabcolsep=0.11cm
\begin{deluxetable*}{llccccccccc} \tablecolumns{11}
    \tablewidth{0pt}
    \tablecaption{GLASS Size--Mass Relations: Best-fit Coefficients}
    \tablehead{\colhead{Region} & \colhead{Type} & \colhead{$\alpha$} &
    \colhead{$\beta$} & \colhead{$\sigma$} & \colhead{$\alpha_1$} & \colhead{$\beta_1$} & \colhead{$\sigma_1$}
      & \colhead{$\alpha_2$} & \colhead{$\beta_2$} & \colhead{$\sigma_2$}\\
      \colhead{} & \colhead{} & \colhead{Single Slope} & \colhead{} & \colhead{} & \colhead{Double Slope} 
      & \colhead{(Low-mass)} & \colhead{}& \colhead{Double Slope} & \colhead{(High-mass)}& \colhead{}
    }
\startdata
Cluster & Passive  & $0.150_{-0.008}^{+0.009}$ & $0.169_{-0.012}^{+0.008}$ & $0.270_{-0.006}^{+0.006}$ & $0.109_{-0.006}^{+0.006}$ & $0.111_{-0.014}^{+0.009}$ & $0.197_{-0.007}^{+0.005}$ & $-0.302_{-0.043}^{+0.045}$ & $0.513_{-0.028}^{+0.032}$ & $0.231_{-0.010}^{+0.014}$\\
Cluster & Star-forming & $0.360_{-0.018}^{+0.014}$ & $0.189_{-0.019}^{+0.019}$ & $0.275_{-0.013}^{+0.011}$ & $0.333_{-0.018}^{+0.018}$ & $0.179_{-0.017}^{+0.018}$ & $0.228_{-0.008}^{+0.008}$ & $0.092_{-0.162}^{+0.120}$ & $0.395_{-0.093}^{+0.136}$ & $0.187_{-0.041}^{+0.020}$\\
Field     & Passive  & $0.168_{-0.014}^{+0.011}$ & $0.150_{-0.016}^{+0.016}$ & $0.224_{-0.013}^{+0.009}$& $0.133_{-0.020}^{+0.021}$ & $0.078_{-0.024}^{+0.018}$ & $0.184_{-0.010}^{+0.009}$ & $-0.067_{-0.133}^{+0.134}$ & $0.340_{-0.091}^{+0.091}$ & $0.213_{-0.031}^{+0.016}$\\
Field     & Star-forming & $0.373_{-0.013}^{+0.010}$ & $0.223_{-0.014}^{+0.012}$ & $0.251_{-0.006}^{+0.004}$ & $0.350_{-0.013}^{+0.011}$ & $0.210_{-0.015}^{+0.014}$ & $0.216_{-0.007}^{+0.004}$ & $0.343_{-0.073}^{+0.082}$ & $0.225_{-0.066}^{+0.063}$ & $0.197_{-0.020}^{+0.019}$
\enddata   
\tablecomments{Summary of the best-fit coefficients for single slope ($\log r_e/\kpc=\alpha + \beta \log (M_*/10^9 \Msun)+N(\sigma)$; see Equation~\ref{eq:canonicalSMR} in the main text) and double-slope fit (same as Equation~\ref{eq:canonicalSMR}, but subscript~1 for $\logm<9.8$ and subscript~2 for $\logm\geq9.8$ galaxies). 
}
\label{tab1}
\end{deluxetable*}


\section{Canonical Size--Mass Relation Analysis}
\label{sec:result1}

While we will ultimately adopt a more sophisticated description (Section \ref{sec:result2}), we begin our analysis of environment's influence on galaxy size and structure by examining ``canonical'' size--mass relations: i.e., by splitting the sample into four populations---passive/star-forming, cluster/field---and comparing linear fits to their $\log r_{e}$--$\log\Mstel$ correlations. We model the samples with a simple regression:
\begin{equation}
	\log\left(\frac{r_e}{\kpc}\right) = \alpha + \beta \log\left(\frac{\Mstel}{10^9\,\Msun}\right) + N(\sigma),
\label{eq:canonicalSMR}
\end{equation}
where $\alpha$ and $\beta$ are the relation's intercept (at $10^{9}\,\Msun$) and slope, and $N(\sigma)$ is a Gaussian describing its intrinsic dispersion assuming that sizes are lognormally distributed at fixed stellar mass \citep[e.g.,][]{newman14}.

We use standard Bayesian techniques to derive the posterior probability of the parameters---including the intrinsic scatter---using a Monte Carlo Markov Chain (MCMC) solver (see Appendix~\ref{sec:mcmc}).

We fit the samples over the entire mass range, but note that $\sigma$ and all parameter uncertainties decrease when the sample is split at $\logm\sim 9.8$ (see Section~\ref{sec:twoPops}), and the low- and high-mass objects fit separately (Table~\ref{tab1}).
This suggests that these populations may be fundamentally different. 
Sections \ref{sec:blueModel} and \ref{sec:discussion} explore this statement further.


\subsection{Size--Mass Relations of Four Populations}
\label{ssec:sizemass}

Figure \ref{fig:MSR} shows the derived linear size--mass relations (Equation~\ref{eq:canonicalSMR}) for the four populations: passive and star-forming galaxies in clusters and the field. 
Outliers excluded from the analysis are plotted as open boxes (see Section \ref{sec:parameters}, Appendix~\ref{sec:Aa}).
The best-fit slopes are derived with each point weighted by its measurement error (see Appendix~\ref{sec:mcmc}).
Figure \ref{fig:slope} and Table \ref{tab1} summarize the results.

For either star-forming or passive galaxies, we find identical slopes and intercepts in both cluster-core and field environments within the uncertainties, suggesting---as previously seen \citep[e.g.,][]{huertas13, vulcani14c}---that environment {\it seemingly} does not affect galaxy size (at fixed mass, population, and time) down to $\logm\sim8$.
Furthermore, while we recover the expected correlations of stellar mass and S\'ersic index \citep[e.g., from][]{lang14} for both sets of galaxies (color-coding in Figure~\ref{fig:MSR}; see also Section~\ref{ssec:sersic}), there is no obvious difference in S\'ersic index trends in clusters or the field, suggesting that environment {\it also} does not affect galaxy structure.

Notably, our derived slopes for passive systems are much shallower than those found previously in either the local universe \citep[$\beta\sim0.6$ by][]{shen03} or at similar redshifts to our sample's \citep[$\beta\sim0.7$ by][fit over $\logm\gtrsim10$]{vanderwel14}.
As detailed in Section \ref{sec:twoPops}, this finding is driven by our low mass-completeness limit---$\logm=7.8$---hitherto unexplored at $z\sim0.5$.


\subsection{Structural Parameters on the Size--Mass Diagram}
\label{ssec:sersic}

Besides size, galaxies' structures can encode the action of evolutionary mechanisms. We now explore this aspect of our systems as parameterized by the S\'ersic index, $n$.

Figure~\ref{fig:sersic} shows the median and $16^{\rm th}$--$84^{\rm th}$ percentile spreads in $n$ for passive/star-forming cluster/field galaxies as a function of stellar mass. 
As was true for their sizes (Section~\ref{ssec:sizemass}), star-forming cluster and field galaxies display almost identical trends, implying that entrance into or life inside the cluster (at late times) induces little if any structural transformation in these systems.

We also see that the S\'ersic index distributions for low mass ($\logm<9.5$) passive and star-forming galaxies overlap significantly in both environments, though passive galaxies---especially in clusters---are offset systematically to slightly higher $n$ values.
Such structural similarities of low-mass star-forming and passive galaxies supports a scenario where low-mass passive systems do not arise through violent mechanisms---such as mergers---but instead gentle phenomena. 
Alternatively (or additionally), they are most consistent with having been drawn exclusively from the high-$n$ tail of the star-forming galaxy population, modulo the effects of disk fading.
We return to these points in Section \ref{sec:discussion}.


\subsection{Two Populations of Passive Cluster Galaxies}
\label{sec:twoPops}

The results above show that galaxies in the highest density regions of the universe do not differ systematically in the size--mass--S\'ersic index plane from those in mean density environments. 
However, this does not mean that clusters do not influence galaxy growth, nor does it mean that they do not trace special regions of the universe. 
The comparisons so far have been gross examinations of four samples at the same epoch. 
Yet, the mechanisms that either are transforming or have transformed the cluster population with respect to that of the field (Figure~\ref{fig:smf}) may act too subtly---or have acted too long ago---to probe in the above fashion. 
Can we find any evidence for this in our data?

To do so, we take two steps. First, we turn our attention exclusively to the low-mass tail of the cluster population: as mentioned in Section \ref{sec:intro}, these systems should be most affected by environment-specific mechanisms (harassment, strangulation, stripping). 
Second, in the next section, we study the sizable ($\simeq0.25\,{\rm dex}$) {\it scatter} of the size--mass relation: it is clearly significant, so perhaps it encodes important information.

Splitting the sample at $\logm \sim 9.8$,\footnote{Shifting the border $\pm0.1\, {\rm dex}$ does not affect the result.} where the slope seems to change, we fit the less- and more-massive galaxies separately with the formula in Equation \ref{eq:canonicalSMR}. 
Figure~\ref{fig:2slopes} shows the results for passive cluster galaxies.

Here, we see that low-mass passive cluster galaxies exhibit a much milder, nearly flat slope of $\beta_1=0.11\pm{0.01}$. 
Put differently, galaxies in this regime increase in size by a factor of $1.7$ for every factor of 100 in stellar mass. 
This is much shallower than either the slope for massive passive galaxies---$\beta_2\sim0.51$ \citep[consistent with previous studies; e.g.,][]{shen03}---or star-forming galaxies of any mass in any environment \citep[$\beta\sim0.2$; e.g.,][]{vanderwel14, allen16, annunziatella16}.
Notably, the inferred intrinsic scatter decreases from $\sigma=0.27\pm{0.01}$ of the single slope fitting to $\sigma_1=0.20\pm{0.01}$ for low-mass, and $\sigma_2=0.23\pm{0.01}$ for high-mass galaxies when these systems are fit separately.
All best-fit results are listed in Table~\ref{tab1}.

These findings qualitatively support previous studies that find the low--mass passive population to have a shallower size--mass relation slope in the local universe \citep[e.g.,][]{binggeli84, ferrarese06, omand14} and at $z\sim0.5$ \citep[e.g.,][]{vanderwel14}.
\citet[][]{ferrarese06} is especially relevant as it is a much more highly resolved \hst\ study of Virgo cluster galaxies. These authors identify a similarly flat trend---$d\log(r_{e})/dM_{B}=-0.05\pm0.02\Rightarrow\beta\approx0.125$ for galaxies with $M_{B}\lesssim-20$ \citep[$\logm\lesssim10.5$, assuming $M/L_{B}$ from][at $B-R=1.4$. See their Equation 21 and Figure 117, top-right]{bell03}. 
Later studies of Virgo dwarf galaxies by \citet[][]{lisker09} and \citet{toloba15} confirm these findings. 
Ultimately, we will concur with \citet{vanderwel14} and \citet{toloba15} in suggesting that some of these objects are likely environmentally stripped low-mass late-types (but also \citealt{lisker09} in that some {\it are not}; Section \ref{sec:discussion}), but, irrespective of any mechanism, a break in this relation suggests the existence of two classes of passive galaxies with distinct formation paths: one with a steep size--mass slope indicative of nearly constant stellar surface density \citep[$\beta=0.5$;][]{hubble26} and therefore perhaps a single formation time, and one with a shallow slope ($\beta_1\approx0.1$) indicative of a broad range of ({\it mostly} lower) stellar surface densities, and therefore a range of formation times. 

We note that the similarly shallow slope is also observed for the passive {\it field} galaxies (Table~\ref{tab1}), where environmental processes are thought to be subtle.
In this case, the shallow slope for low-mass galaxies would not be attributed to the environmental effect.
However, low-mass passive field galaxies in this study could be in similar environments with cluster member galaxies, because we have grouped the sample into the two subgroups by using the same FoV.
Actually, \citet{geha12} found that most of low-mass passive galaxies are in group (or denser) environments by using the local sample, and so would be affected by processes similar to those affecting the low-mass cluster sample.
Our {\it genuine} field sample, which resides in PR1 and XDF regions, is statistically weak and larger data sets will be required to fully address this question.


\subsection{Toward a Formation Pathway:\\ Star-forming Galaxies as a Model for Passive Galaxies}
\label{sec:blueModel}

In any scenario, passive galaxies were star-forming at some epoch.
Hence, we can use the star-forming galaxy size--mass relation derived above as a model for the sizes that passive galaxies {\it should} have if they had stayed in the star-forming population. 
Residuals from this exercise---the difference between how big the passive galaxies are and how big they are predicted to be---may contain important information about how, when, and therefore why they diverged from their star-forming peers. 

Figure~\ref{fig:2slopes} compares the size--mass relation of passive cluster galaxies with the mean relation for star-forming galaxies at several redshifts taken from \citet{vanderwel14}. 
Points are color-coded by the rest-frame $U - V$ color.
This quantity is a good indicator of the time since a galaxy's last episode of star formation (at least out to $\sim 4\, {\rm  Gyr}$ for low-mass objects, assuming the local stellar mass--metallicity relation of \citealt{kirby13} holds at $z\sim0.5$) and therefore serves as a clock counting back to when any red galaxy was last in the star-forming population.
Evidently, only the largest of the passive low-mass systems have sizes compatible with equal mass star-forming (field) galaxies near the epoch of observation ($0.2\leq z\leq0.7$), and are therefore consistent with being drawn from that population.
The sizes of some very high mass passive systems are also comparable with those of star-forming galaxies at the same epoch, but since the former are giant ellipticals, effectively all of their other properties rule them out from having descended from the latter.

Figure~\ref{fig:lowQG} explores these findings in greater detail, showing the size {\it difference} of passive cluster galaxies from the star-forming relation at $z\sim0.5$ \citep[taken from][]{vanderwel14}.
Boxes highlight the median $U-V$ colors in $0.2\times0.1\,{\rm dex}$ boxes.
The horizontal lines also shown are the predicted offsets derived from the size-mass relations at different redshifts also from \citet{vanderwel14}.

A concern here is that, as seen in the left panel of Figure \ref{fig:MSR}, star-forming galaxies exhibit a color--mass trend, such that more massive star-forming objects are redder. 
Hence, the observed color gradients in Figure \ref{fig:lowQG} might reflect a baseline mass--color covariance in the star-forming (source) population, not a true third parameter.

To correct for this, we fit the mass--color relation for star-forming galaxies, obtaining:
\begin{equation}\label{EqUVcor}
	U-V{\rm /mag} = 1.03 + 0.11\,\log\left(\frac{\Mstel}{10^9\, \Msun}\right).
\end{equation}
As shown in the right panel of Figure~\ref{fig:lowQG}, even after applying this correction to the passive cluster galaxies, the color gradient along the $y$-axis remains, confirming our earlier statement that, to some degree, the spread of passive cluster galaxy sizes at fixed mass reflects a record of {\it the time} when a system left the star-forming population. 
As this extends to at least $z\sim3$ (i.e. $\sim7$ Gyr before the epoch of observation), this gradient can encode quite long timescales \citep[][]{speagle14, abramson16}.

A clear, quantitatively robust color gradient is apparent at fixed stellar mass, shown in Figure~\ref{fig:lowQG2}.
As clarified by the colored squares in Figure~\ref{fig:lowQG}, smaller-size passive galaxies are also {\it redder} than larger-size passive galaxies. 
This finding amplifies results from Figure \ref{fig:2slopes}: at low masses, the largest-size passive galaxies have both sizes {\it and} colors consistent with their having been been drawn from the star-forming population more recently than their smaller-size passive counterparts.

Taken together, Figures \ref{fig:2slopes}--\ref{fig:lowQG2} present strong evidence that smaller passive galaxies ``quenched'' earlier, while the largest passive galaxies are quenching now.
The same trend is observed in local clusters \citep[e.g.,][for $\logm>9.8$ galaxies]{valentinuzzi10} and in the field \citep[e.g.,][$\logm>10.3$]{poggianti13}, with luminosity-weighted age, rather than $U-V$ color, which support our interpretation.

Interestingly, our analysis shows no such gradient for massive passive galaxies ($\logm>9.8$; Figure~\ref{fig:lowQG2}), supporting the idea that they have a different formation history than many low-mass objects, at least in the sense of having been in place long before the epoch of observation.
This conclusion is also supported by the nearly uniform stellar surface densities of these objects (Figure \ref{fig:2slopes}; $\beta_2\sim0.5$ corresponds to $M_*/r_e^2={\rm const}.$), compared to the large spread of densities in the low-mass population.
This homogeneity of massive passive galaxies is also found by \citet{zanella16} with $\hst$ spectroscopic measurements at $z\sim2$.

These facts---combined with the detailed properties of the massive galaxies' stellar populations \citep{thomas05,mcdermid15}---suggest an {\it accelerated} growth channel for the high mass population, rather than environmental quenching, which may dominate at low masses.
The low-mass population is converged to a high-mass one at low $\Delta \log r_e$.
This may indicate that these low-mass galaxies trace the same channel as the high-mass galaxies.
We return to this statement in Section \ref{sec:discussion}.

Our ability to identify and compare this trend with non-cluster low-mass passive galaxies is hampered by the very small sample size. 
The same exercise reveals a trend in the same direction as in the cluster sample, but with large uncertainties. 
The mean color of the field systems is bluer than their cluster counterparts at these low masses, another point we will return to in Section \ref{sec:discussion}.

\tabletypesize{\footnotesize} \tabcolsep=0.11cm
\begin{deluxetable*}{ccccccc} \tablecolumns{7}
    \tablewidth{0pt}
    \tablecaption{GLASS Size--Mass Relations: Best-Fit Coefficients for the holistic fitting (Eq.~\ref{eq:bigEq} and Figure~\ref{fig:NFIT}).}
    \tablehead{\colhead{$\alpha$} & \colhead{$\beta_{M_*}$} & \colhead{$\beta_{n}$} & \colhead{$\beta_{UV}$} & \colhead{$\beta_{\Sigma_5}$} & \colhead{$\beta_{z}$} & \colhead{$\sigma$}\\
    }
\startdata
$0.122_{-0.005}^{+0.005}$ & $0.254_{-0.008}^{+0.007}$ & $-0.041_{-0.020}^{+0.019}$ & $-0.286_{-0.014}^{+0.014}$ & $-0.031_{-0.007}^{+0.006}$ & $-1.059_{-0.130}^{+0.142}$ & $0.205_{-0.003}^{+0.003}$
\enddata   
\label{tab2}
\end{deluxetable*}

\begin{figure*}[t!]
\begin{center}
\includegraphics[width=0.9\linewidth]{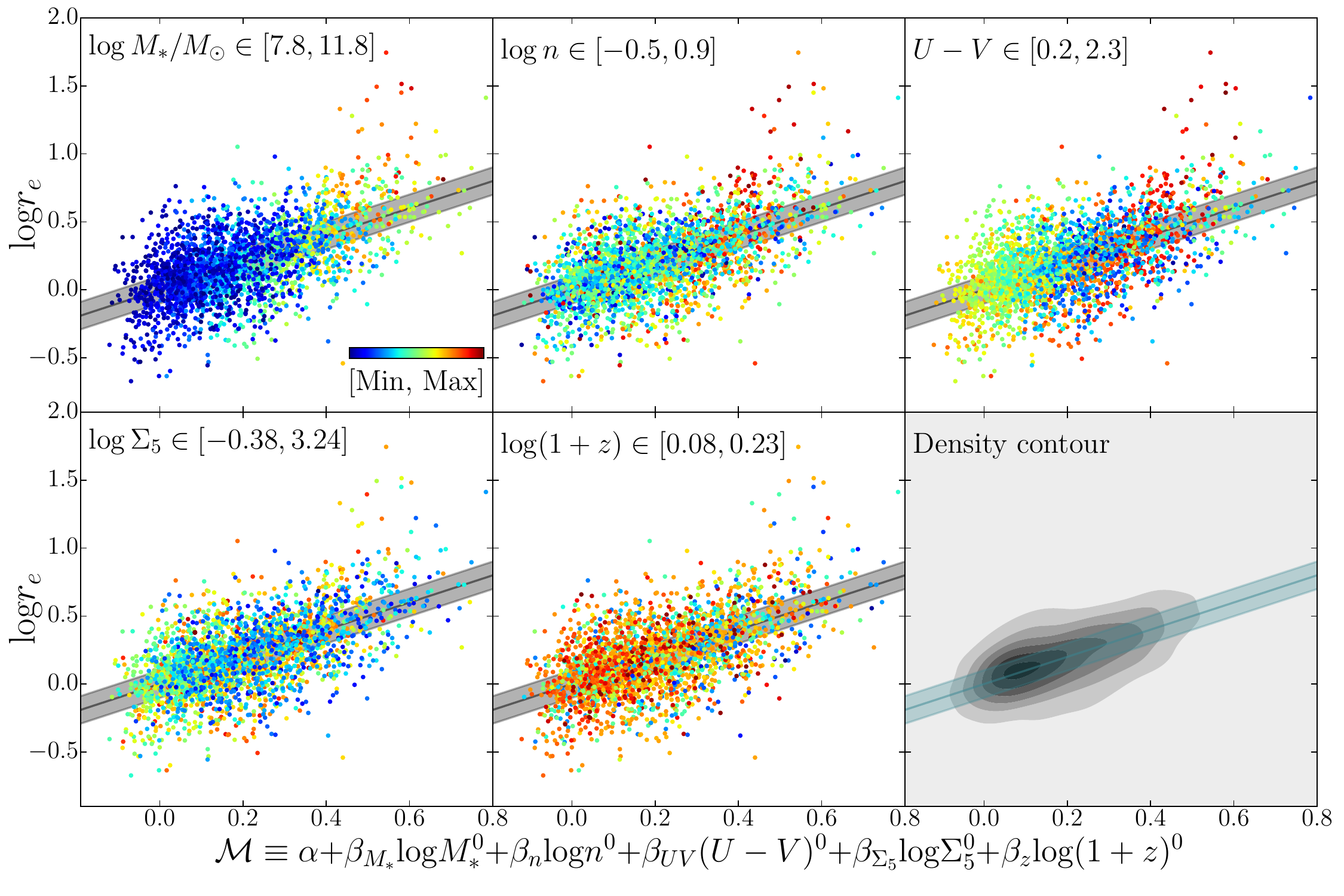}
\caption{
	Galaxy size fitted with stellar mass, S\'ersic index, $U-V$ color, local density ($\Sigma_5$), and redshift (see Equation \ref{eq:bigEq}).	
	In this fitting, galaxies are not split into any sub-categories (e.g., cluster/field, red/blue). 
	$^{0}$ superscripts in the equation denote that the variables have been normalized to the pivot values (medians) given below Equation \ref{eq:bigEq}.
	We show the same best fit result in 6~panels, but each of which has different color-coded scheme : stellar mass (top left), $\log n$ (top middle), $(U-V)$ (top right), $\log \Sigma_5$ (bottom left), and $\log(1+z)$ (bottom middle). 
	The contour of galaxy number density is shown in bottom right.
	The color is coded according to the observed parameter range shown in each bracket ([Min,Max]).
	}
\label{fig:NFIT}
\end{center}
\end{figure*}

\begin{figure*}
\begin{center}
\includegraphics[width=0.8\linewidth,bb=0 0 576 576]{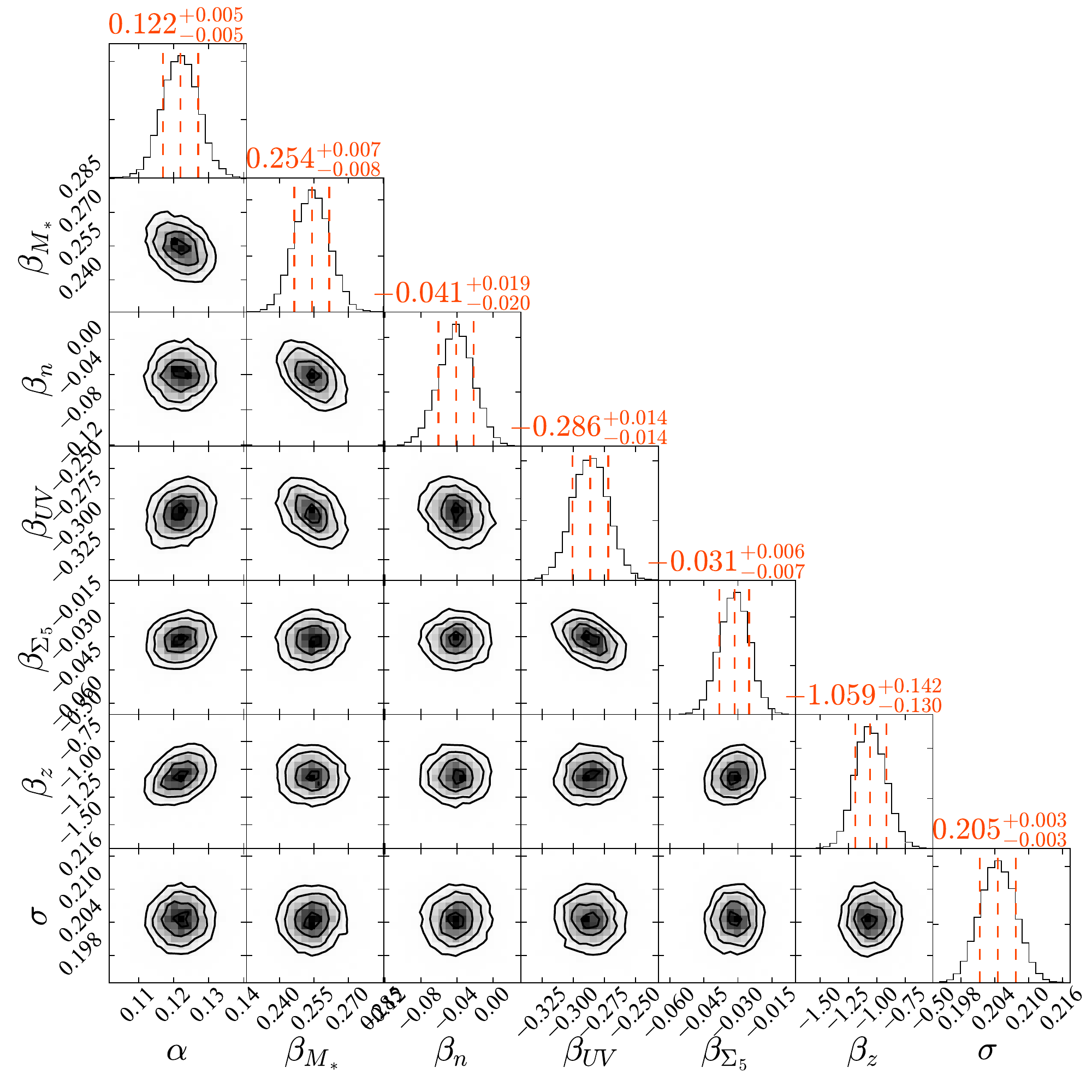}
\caption{
	Parameter estimates for the ``holistic size--mass relation'' (Equation \ref{eq:bigEq}) fit to the full galaxy sample.
	Contours reflect 68\%, 96\%, and 99\% confidence intervals as determined using an MCMC solver.
	The best fit coefficient values, 50th~percentile, are shown on top of each column with offsets from 16th/84th~percentiles.
	We see no significant degeneracies between the derived coefficients.
	}
\label{fig:mcmc}
\end{center}
\end{figure*}


\section{A Holistic analysis of all galaxies in higher-dimensional parameter space}
\label{sec:result2}

So far, we have performed a canonical examination of the size--mass relation (Section \ref{sec:result1}).
When separating samples by color and environment {\it a priori}, it seems that the relation is not so sensitive to where a galaxy is, but rather {\it when} it left the star-forming population (Figure \ref{fig:lowQG}).

Technically, however, the above comparisons are not the fairest tests of environmental effects. These require matching cluster and field samples in all relevant characteristics---e.g., color, mass, and S\'ersic index---leaving environment as the {\it only} distinguishing trait. 
Here, we adopt a multidimensional analysis that takes the above factors into account, and thus provides a fair, quantitative assessment of environmental effects.

Combining all galaxies into a single sample, we change our description from one in which mass is the only independent variable to one where a range of other parameters might also influence galaxy size, including: S\'ersic index, $U-V$ color, redshift, {\it and} local number density (i.e., environment). 
Rather than imposing somewhat artificial boundaries, this procedure allows for a data-driven exploration of the correlations between continuous parameters.
The new multidimensional correlation becomes:

\begin{equation}
	\log\left(\frac{r_e}{\kpc}\right) = \mathcal{M} + {\it N}(\sigma),
\label{eq:bigEq}
\end{equation}
where:
\begin{align*}
	\mathcal{M} \equiv \alpha &+ \beta_{\Mstel}\log\left(\frac{\Mstel}{10^{9}\,\Msun}\right)\\
		& + \beta_{n}\log\left(\frac{n}{1.50}\right) \\
		& + \beta_{\Sigma_5} \log\left(\frac{\Sigma_{5}}{214.0\, {\rm Mpc^{-2}}}\right)\\
		& + \beta_{z}\log\left(\frac{1+z}{1.54}\right)\\
		& + \beta_{UV} [(U-V)-1.43\, {\rm mag}].
\label{eq:mDef}
\end{align*}
Here, $\Sigma_{5}=(5+1)/{\pi r_{5}^2}$ is the projected number density (per sq.\ Mpc) defined over the distance to the $5^{\rm th}$ closest object, and all variables are fit with respect to the pivot values (medians) in their respective denominators. Figure~\ref{fig:NFIT} shows the best-fit relation, whose parameters are listed in Figure~\ref{fig:mcmc} and Table~\ref{tab2}.

The derived coefficients reveal how much galaxy sizes depend on one parameter when the others are held fixed; they effectively describe how samples differ in one property when matched in all others. 

From these results, we see that stellar mass and color have the strongest correlations with galaxy size: when $\logm$ increases by $1.0\,{\rm dex}$, $\log r_{e}/\kpc$ changes by $\beta_{M_*} \times 1.0\sim0.25\,{\rm dex}$; nearly a factor of 2. Similarly, a $1.0\,{\rm mag}$ increase in $(U-V)$ color results in $\sim0.29\,{\rm dex}$ {\it decrease} in size, consistent with the results in Figure \ref{fig:lowQG}.\footnote{The largest {\it absolute} coefficient corresponds to redshift effects---$\beta_{z}=-1.06\,{\rm dex}/{\rm dex}$---but, the small $z$ range observed ensures this does not translate to a large impact on measured sizes: a $0.3\,{\rm dex}$ ($2\times$) change in $r_{e}$ via redshift evolution alone requires comparing $z\sim2$ systems to our sample (consistent with Figure \ref{fig:lowQG}).}

In this context, the nearly complete independence of galaxy sizes on environmental density deduced from Figure \ref{fig:slope} is made dramatically and quantitatively clear. 
The best-fit coefficient is $\beta_{\Sigma_5}=-0.031$, corresponding to only $\Delta\log r_{e}/\kpc<0.1$ dex ($<26\%$) over the factor of $\sim 1000$ spread in projected density probed by our data ($1\lesssim\Sigma_5/\,{\rm Mpc}^{-2}\lesssim1000$).
We stress again that our sample spans normal environments to the densest regions in the universe---cluster cores. 
Hence, it seems unlikely that a stronger signal could be found by looking elsewhere.

We have confirmed that the residuals produced by applying the globally fit holistic model to the field and cluster subsamples separately are flat, and have a dispersion consistent with that shown in Figure \ref{fig:NFIT}. Hence, cluster galaxies with low $\Sigma_{5}$ (e.g., those at large $R_{\rm cl}$) and non-cluster galaxies with high $\Sigma_{5}$ (e.g., those in groups), do not appear to deviate significantly from expectations derived without knowledge of the global environment.

While these results are consistent with our previous findings
based on a simpler analysis, the treatment described here presents
several advantages. 
First, it allows the {\it data} to identify which correlations are most important, rather than humans using bins based on what we think {\it should} be relevant.
Second, by making all variables explicit, we avoid being misdirected by correlations imposed by hidden parameters. 

Indeed, had such an analysis been favored {\it a priori}, it would have been immediately evident that any analysis of the effect of environment {\it at fixed epoch} would provide only partial answers. 
As shown in Figure \ref{fig:lowQG}, most of the well known ``environmental'' trends are in fact a reflection of differences in galaxy {\it ages}, and perhaps ancient discrepancies in the mass functions that differentiate the structures that collapsed first in the universe (clusters) and those that do so much later (the field; see Figure \ref{fig:smf}, also \citealt{kelson16}). 
Beyond this, the gross appearance of such scaling relations is apparently highly insensitive to any cluster-specific mechanisms.

Combined with the residual color-dependence---a ``clock'' measuring the time since a passive galaxy left its star-forming peers---this finding reinforces suggestions from the previous analyses that local number density {\it traces} transformative phenomena to an important extent.
That is, it marks regions wherein an initial large-scale overdensity caused all systems within it to evolve rapidly, independent of any late-time {\it transformative} effects. 
We discuss this further below.


\section{Discussion}
\label{sec:discussion}

So far, we have studied galaxy size and structure in different environments by (1) splitting the sample into four sub-populations (the ``canonical'' approach; Section \ref{sec:result1}), and (2) treating them as a single population (the ``holistic'' approach; Section \ref{sec:result2}).

Via the first approach, we see no environmental effect on the size--mass relation, even at the unexplored stellar mass limit of $\logm=7.8$ at $z\sim0.5$.
Via the second method, this finding is qualitatively confirmed, and quantitatively contextualized: assuming only that galaxies can be described by a suite of parameters that should be sensitive to the same phenomena impacting galaxy sizes, we find that environment has perhaps the {\it smallest} effect.

Figures \ref{fig:2slopes}--\ref{fig:NFIT} suggest that, for many systems, by the time a galaxy is sufficiently ``transformed''---by whatever process---to be identified as ``passive,'' evidence of any direct impact of cluster-specific phenomena is wiped out, or of secondary importance at best.

This seems certainly true for high-mass cluster galaxies ($\logm\gtrsim9.8$), which dominate the cluster's stellar mass content and have stellar populations that are completely incompatible with having been drawn recently from the star-forming population.
For low-mass cluster galaxies, however, the picture is more nuanced. 

As discussed in Section \ref{sec:blueModel} (Figures \ref{fig:lowQG} and \ref{fig:lowQG2}), the largest-$r_{e}$ passive cluster galaxies with $\logm<9.8$ have sizes and colors consistent with their having come from the field star-forming population at times close to the epoch of observation. 
The smallest-$r_{e}$ galaxies in this population, however, seem to have been in place for many Gyr, perhaps as long as their higher-mass neighbors, suggesting a dual formation scenario for low-mass passive galaxies \citep[see also, e.g.,][]{poggianti06}.


\subsection{Large-$r_{e}$ Low-mass Passive Cluster Galaxies}
\label{ssec:llmass}

On the face of it, the fact that the largest-size low-mass passive galaxies lie near the size--mass relation of star-forming galaxies at the same epoch points immediately to something like ram pressure stripping or starvation \citep[e.g.,][]{wetzel13} as the most likely transformative mechanisms.
These are indeed cluster-{\it driven}, relying exclusively on the properties of the mature cluster environment. 
Four additional facts support this conclusion: 
\begin{enumerate}
	\item Our target clusters are {\it currently} bright X-ray sources, confirming the presence of dense intra-cluster gas \citep{mantz10}. 
	Especially in the core regions probed by the \hst\ observations, drag from this hot atmosphere can effectively strip gas from infalling galaxies, and also stifle their accretion of new gas for future star formation, the definitions of ram-pressure stripping and starvation.
	\item Large-size, low-mass systems are most amenable to (ram pressure) stripping as their internal gas supplies are most loosely bound \citep{treu03}. They would also run out of fuel for star formation relatively quickly if starved of external fuel supplies given the generally higher specific star formation rates of their low-mass star-forming galaxy progenitors \citep[e.g.,][]{salim07, whitaker14}.
	\item The low-mass end of the passive cluster mass function grows over the epochs probed (Figures \ref{fig:smf}, bottom-right), consistent with a scenario where star-forming galaxies are continually being transformed.
	This is also consistent with previous studies that a single epoch is disfavored for the formation of low-mass passive cluster galaxies \citep[e.g.,][]{roediger11,toloba14}.
	\item The relatively low S\'ersic indices of the low-mass passive cluster galaxies (Figure \ref{fig:sersic}) point to a ``gentle'' mechanism; one that does not destroy disks/rearrange galaxies' stellar components, which is consistent with starvation, ram pressure stripping, and galaxy harassment \citep{bialas15}. 
	The observed S\'ersic indices of low-mass passive galaxies are slightly higher than those of star-forming ones with similar stellar mass, but this difference is explained by disk fading, where cessation of star formation in the disk leads to more concentrated light profiles (and thus higher S\'ersic indices; \citealt{lackner13}).
\end{enumerate}
Given the direct evidence for stripping in local analogues \citep[e.g.,][]{cayatte90,abramson11} and at intermediate redshift \citep{vulcani15}, it seems likely that at least this mechanism is {\it currently} operative.

{\it However}, to attribute the presence of the smallest-size low-mass passive galaxies---which are {\it also} the oldest (Figure \ref{fig:lowQG})---to the same mechanism(s), one must take into account the evolution of the cluster itself. 


\subsection{Small-$r_{e}$ Low-mass Passive Cluster Galaxies}
\label{ssec:slmass}

At $z\sim3$, when many of these $\logm<9.8$ systems seem last to have been in the star-forming population, our target clusters would have been much less massive, and were therefore home to much more tenuous, cooler intra-cluster media.
Based on calculations by \citet{trenti08},\footnote{$\Lambda$CDM cosmology based on Wilkinson Microwave Anisotropy Probe (WMAP) year 1 results \citep{spergel03}.} we estimate that our clusters---systems with $\log M_{\rm halo}/\Msun\sim15$ at $z\sim0.5$---had progenitors with $\log M_{\rm halo}/\Msun\sim14$ at $z\sim3$ (also consistent with \citealt{evrard02}). 
Hence, the question becomes whether or not the environments at those epochs---when the global density and neutral gas fraction was higher---could have supported ram pressure stripping/significantly cut galaxies off from their fuel supplies. 
If they could, these channels could provide a unified explanation for all low-mass passive cluster galaxies. 
If they could {\it not}, another channel must have been open.

\begin{figure}
\begin{center}
\includegraphics[width=8.2cm,bb=0 0 288 288]{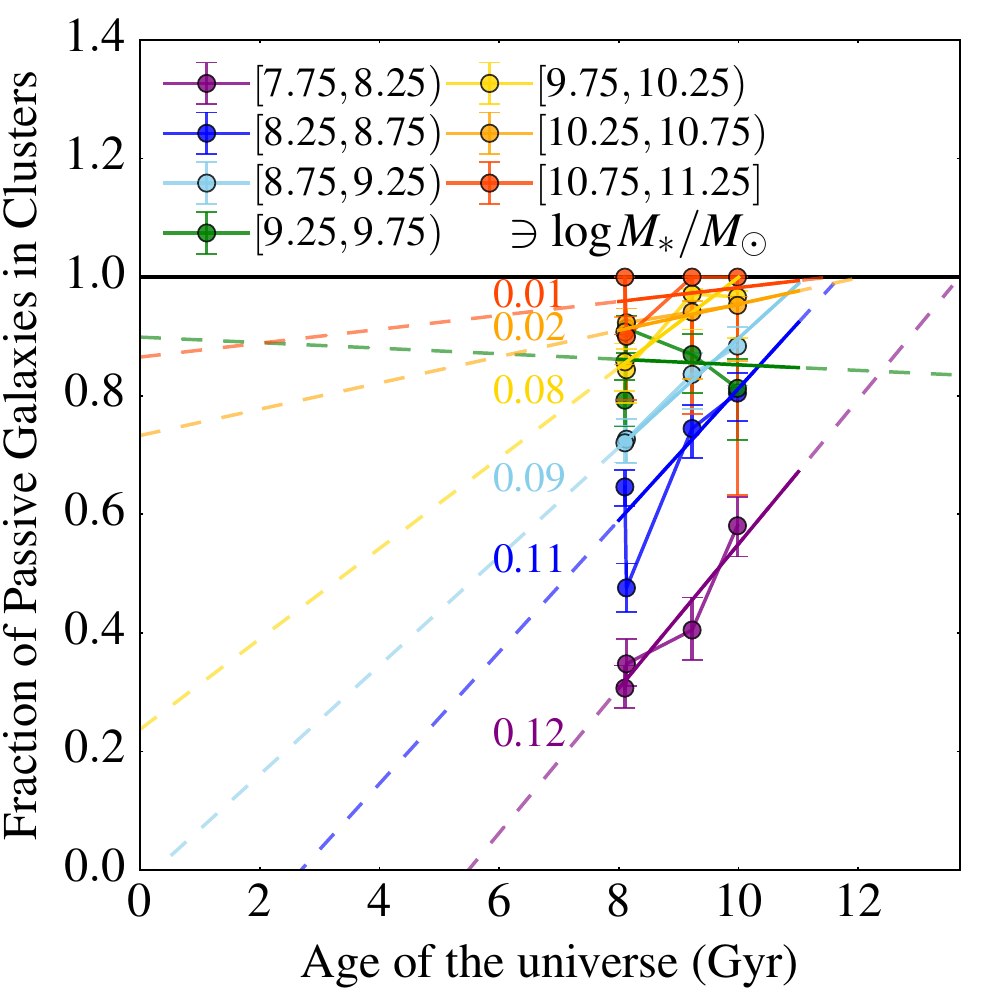}
	\caption{Extrapolation of the current quenching rate to high lookback times as a function of stellar mass. Each color corresponds to stellar mass bin listed in legend.
	The derived linear fit slope, $dF_{\rm red}/dt$, for each stellar mass bin is shown near the fitting line in the same color.
	One stellar mass bin, $9.25<\logm<9.75$ (green), has a negative slope, $dF_{\rm red}/dt=-0.004$, because of a weak statistics, and the value is not shown.
}
\label{fig:frac_evl}
\end{center}
\end{figure}

To further constrain the strength of the above channels and their ability to produce the smallest low-mass passive cluster galaxies, we can assume all of the evolution of the passive fraction shown in Figure \ref{fig:smf}, bottom-right, is due to the same mechanism(s) and project the effects back in time. 
We do so by extrapolating a simple linear regression and show the results in Figure \ref{fig:frac_evl}.
This is an extreme model---we might expect the recent evolution in passive fractions to be more rapid than its past evolution due to the cluster and cosmic evolutionary effects mentioned above, which should reduce the efficiency of, e.g., stripping, with lookback time.
However, it provides something like an upper limit to the quenching that could be driven by such mechanisms, which is what we seek.

From this exercise, we obtain $dF_{\rm red}/dt\sim0.1\, {\rm Gyr^{-1}}$ at $\logm<9$.
By extrapolating the slope, we find that the passive fraction at these masses would be zero at $2\lesssim z\lesssim4$, depending on the bin.
Hence, it could be that {\it all} cluster (core) galaxies with $\logm<9$ were transformed due to gas removal, or whatever other process is active {\it now} in clusters.

However, this idea does not hold for even slightly more massive passive cluster galaxies; i.e., those with $9.25\lesssim\logm\lesssim10.25$---there are simply too many of these to have all arisen through the same channel.
Our toy calculation suggests that perhaps $30\%$ of these systems were already in place at $z\sim3$, just 2~Gyr after the Big Bang.
For these galaxies---and presumably those with yet lower masses, assuming a more-realistic nonlinear $dF_{\rm red}/dt$---other explanations must be sought. 
We explore one possibility below.


\subsubsection{Evidence for Accelerated Evolution for Low-mass Dense Cluster Galaxies}

Clues for a formation scenario come from the low-mass passive {\it field} and the high-mass cluster populations.

From Figure \ref{fig:uvj}, we see that, at $\logm<9$, there exist passive galaxies in clusters that are systematically redder than those in the field. 
This implies that the cluster galaxies {\it reached their final mass before their field counterparts}.

Turning to the $\logm>9.8$ passive cluster population, Figure \ref{fig:2slopes} shows the size--mass relation of these objects to lie remarkably close to a line of constant stellar surface density, $\Sigma_*=\Mstel/r_{e}^{2}$. This points to common formation time for these systems \citep[][]{franx08, vandenbosch08, stringer14, lilly16, whitaker16, abramson16b}.
Given their uniformly ancient stellar populations (based on their colors), the implication is that such high-mass passive cluster galaxies are monolithically old and do not descend from field galaxies transformed over long stretches of time.

Assuming this is the case, we can obtain a rough idea of how many low-mass objects formed similarly by calculating the fraction of them that have surface densities similar to the high-mass systems. 
Given that the 2-$\sigma\approx0.5\,{\rm dex}$ intrinsic spread in sizes at fixed mass corresponds to a factor of 100 difference in surface densities for galaxies at the top and bottom of the size--mass relation, we should expect some $\logm=8$--9 passive objects to have densities comparable to their $\logm=10$--11 counterparts.

Indeed, we find $\approx18\%$ of galaxies with $\logm<9.8$ to lie above $\Sigma_*\approx10^{8.3}\, \Msun\,\kpc^{-2}$---the 1-$\sigma$ lower bound to the high-mass sample's surface densities. 
This fraction is close to that independently obtained by extrapolating $dF_{\rm red}/dt$ above. 
Combined with the fact that these systems will, by definition, be the smallest low-mass galaxies---and therefore also the {\it reddest} (Figure \ref{fig:lowQG})---this finding strengthens the conclusion that such systems were in place long ago, having formed alongside their massive counterparts.
This is consistent with \citet{toloba15}, who find low-mass galaxies with higher stellar velocity dispersions at fixed mass to have lower $r_{e}$ and lie closer to the cluster core in Virgo, and therefore be older than larger-$r_{e}$ systems.

Notably, the same density calculation reveals only $6\%$ of low-mass {\it field} passive galaxies to have densities consistent with high-mass passive objects. 
If we assume, following \citet{geha12}, that all of these dense systems are in fact stripped satellites---i.e., they do not truly arise from the same processes generating high-mass passive galaxies---this estimate can be taken as our measurement uncertainty.

As such, combined with the $dF_{\rm red}/dt$ results, we can state the following: regarding the formation of most high-mass and 10\%--30\% of $\logm<9.8$ passive cluster galaxies, clusters mark regions of space where evolution was {\it accelerated} due to a population's residence in a common overdensity.
That is, perhaps {\it all} sufficiently dense cluster galaxies---regardless of stellar mass---are consistent with having arisen through a common, prompt, formation channel.
This scenario---consistent with that of \citet{dressler80}, \citet{abramson16}, and \citet{kelson16}---is fundamentally different from the environmental quenching of infalling field systems that accounts for the rest of the (mainly low-mass) cluster passive galaxy population at $z\sim0.5$, and presumably today,
suggesting a dual (or bimodal) formation scenario for passive cluster galaxies \citep[see also][]{poggianti01,poggianti06}.

A consequence of any ``accelerated'' growth is that passive cluster objects would lock-in the smaller sizes of {\it all} star-forming galaxies {\it everywhere} at early epochs, appearing naturally at the bottom of the size distribution for their ``final mass'' and exhibiting globally higher S\'ersic indices, precisely as seen in Figures \ref{fig:sersic} and \ref{fig:lowQG}.
This effect would also naturally lead to the color--size anticorrelation revealed by our holistic fit ($\beta_{UV}=-0.29\pm0.01$; Figure \ref{fig:NFIT}), which is therefore actually a reflection of the well-known redshift--size anticorrelation \citep[e.g.,][]{newman12, newman14, vanderwel14, morishita15} that our analysis also reveals ($\beta_{z} = -1.06\pm0.14$).
The evidence of the accelerated grown in dense environment is also observed in a higher redshift cluster \citep[e.g.,][though limited to massive galaxies]{papovich12, bassett13}.

In sum, our results point to an identifiable ``native'' population of galaxies at all masses that matured rapidly at early times {\it because} it was situated in a region of space that was also collapsing quickly. 
This is the mechanism that reaches to large distances, causing the population differences between clusters and the field to extend to many virial radii (\citealt{lewis02}; \citealt{treu03}; \citealt{dressler13}). To this is added a frosting of new galaxies at late times driven by something akin to ram pressure stripping or starvation (certainly something gentle), which is especially active at small clusto-centric radii and later epochs.


\subsection{Better Tests than Scaling Relations}

The HFF provides something close to the best possible imaging data acquirable for objects in the distant universe.
As such, it is unclear what more-detailed studies of the size--mass relation will uncover in terms of offsets between mean sizes of populations as a function of environment that cannot be gleaned already.

Our approach in Section \ref{sec:result2} points to the fundamental limitations of ever more sophisticated analysis of these kinds of galaxy-integrated, photometric metrics. Instead, our analysis suggests that, if one seeks better knowledge of the detailed physical mechanisms transforming galaxies in clusters (or outside of them), different kinds of data are required. Principally, the addition of {\it spectroscopic} data, and probably in a spatially resolved sense; i.e., deep and wide IFU surveys investigating the star formation and kinematic properties that may differentiate galaxies as a function of mass and environment. 

For example, using light-weighted stellar ages, $\logm>9.8$ passive galaxies observed in local clusters and the field have been seen to show a similar trend to our $\logm<9.8$ systems, such that the largest-$r_{e}$ galaxies are the youngest \citep[e.g.,][]{valentinuzzi10, poggianti13}. 
At $z\sim0.5$, we see no significant color--size trend for equal-mass galaxies in this regime, but $U-V$ colors are limited as an age indicator: any passive galaxies with ages $>2$~Gyr would be uniformly red in this index. 
Hence, deep optical spectroscopy of our sample is needed to determine whether any real evolution at these masses takes place in the $\sim5$ intervening Gyr.

Also, the observed scatter in $r_{e}$ could be due to radial stellar migration and not age.
Induced by stellar feedback, such migration can cause $r_{e}$ to fluctuate by $\sim2\times$ in just $\sim100$~Myr \citep[e.g.,][]{elbadry16}. 
We see a clear correlation between color and galaxy size, which suggests such rapid effects are not the principal source of scatter in $r_{e}(\Mstel)$ in the low-mass population, but resolved spectroscopy would constitute a much more stringent physical constraint.

The GLASS \citep{jones15, treu15, vulcani15, vulcani16a, vulcani16b}, GASP (Poggianti et al. in preparation), {\sc Sauron} \citep{davies01}, {\sc Atlas3D} \citep{cappellari11}, {\sc 3D-HST} \citep{brammer12,nelson15}, CALIFA \citep{sanchez12}, {\sc MaNGA} \citep{bundy15}, SAMI \citep{allen15}, and KROSS \citep{magdis16} have already started these investigations, and highly resolved analyses (mainly at low-$z$) show both signs of accelerated evolution driven by clusters \citep{mcdermid15}, and stripping \citep{conselice01,conselice03a,nipoti03,janz16}.
By studying the sites of star formation (\citealt{wang15,wang16}) and kinematics \citep[e.g., KLASS;][]{mason16} through resolved gas maps at high(er)-$z$ using the next generation of ground- and space-based instruments, we may be able to map the evolving importance of these effects at levels of detail currently available only in the local universe.


\section{Summary}
\label{sec:summary}

We derived photometric redshifts, stellar masses, and structural parameters for $>3900$ cluster and field galaxies at $0.2\leq z\leq0.7$ from the HFF and GLASS programs, complete to $\logm=7.8$---an unexplored regime at these redshifts. Using this homogeneous sample:

\begin{enumerate}

	\item We studied the size--mass relations of four ``canonical'' populations---cluster/field, passive/star-forming galaxies---fitting each subsample with a single slope. 
	Though the populations reside in environments of maximally different density, $\sim$ a factor of 1000, we find the relations to be identical within their measurement uncertainties (Figure \ref{fig:slope}).
	This holds even at the lowest masses where cluster-specific effects would be expected to have the most significant impact on galaxy structure.
	\item A multivariate analysis---wherein all galaxy classifications are removed and sizes are fit as a function of stellar mass, S\'ersic index, color, redshift {\it and} environment---is consistent with the above results, and quantitatively reveals local density to induce but a $7\%\pm3\%$ reduction in size ($95\%$ confidence) when controlling for these other factors (Figure \ref{fig:NFIT}). 
	Immediate environment therefore appears to have a tiny effect on galaxy size, while stellar mass and color correlate most strongly.
	\item We studied the trends in $(U-V)$ color in the low-mass passive cluster population ($\logm<9.8$) as a function of offset from the best-fit slope of {\it star-forming} galaxies at the same redshift. 
	We find that smaller-size galaxies are also redder. 
	\item The {\it largest}-size low-mass passive cluster galaxies---which are also the bluest---have sizes and S\'ersic indices similar to those of contemporaneous star-forming galaxies (Figures~\ref{fig:sersic}, \ref{fig:2slopes}).
	This fact suggests that they are recently acquired systems that have been ``quenched'' by a cluster-specific process that terminates star formation in a non-violent manner/preserves the structure of star-forming galaxies.
	Given that our clusters all harbor dense intra-cluster gas, the most likely candidate is ram pressure stripping or starvation.
	\item This explanation holds for the {\it smallest} low-mass passive cluster galaxies only if the progenitors of our clusters were capable of hosting a hot intra-cluster medium at $z\gtrsim3$.
	If not, the consistent stellar surface densities and colors of these objects with those of their uniformly ancient, more massive ($\logm>9.8$) peers suggest that $10\%$--$30\%$ of these galaxies are ``native,'' having had their evolution accelerated---not terminated---by their presence in a large overdensity at birth. 

\end{enumerate}

Our conclusion is therefore not that environment has no impact on galaxy evolution, but rather that it {\it encodes} the fact that most high-mass and $\sim18\%$ of low-mass passive galaxies found in dense regions at late times had common (accelerated) evolutionary trajectories, with late-time effects playing some role, but only dominant at low masses.

\acknowledgements

The authors thank an anonymous referee for carefully reading our manuscript and a number of insightful suggestions.
The authors thank Anton Kokoemoer and HFF team for producing and making publicly available their data. 
We are very grateful to the staff of the Space Telescope for their assistance in planning, scheduling and executing the observations.
Support for GLASS (HST-GO-13459) was provided by NASA through a grant from the Space Telescope Science Institute, which is operated by the Association of Universities for Research in Astronomy, Inc., under NASA contract NAS 5-26555.
Support for this work is provided by NASA through a Spitzer award issued by JPL/Caltech, HST-AR-13235 and HST-GO-13177. 
This study made use of a number of programs provided by the {\it Astropy} software community \citep{muna16}.
T.M. acknowledges support from the Japan Society for the Promotion of Science (JSPS) through JSPS research fellowships for Young Scientists.
B.V. acknowledges the support from an Australian Research Council Discovery Early Career Researcher Award (PD0028506).
A.H. acknowledges support by NASA Headquarters under the NASA Earth and Space Science Fellowship Program - Grant ASTRO14F-0007.


\section*{\small Appendix A\\ICL Subtraction}
\label{sec:Aa}

\begin{figure*}
\centering
	\includegraphics[width=14cm,bb=0 0 547 288]{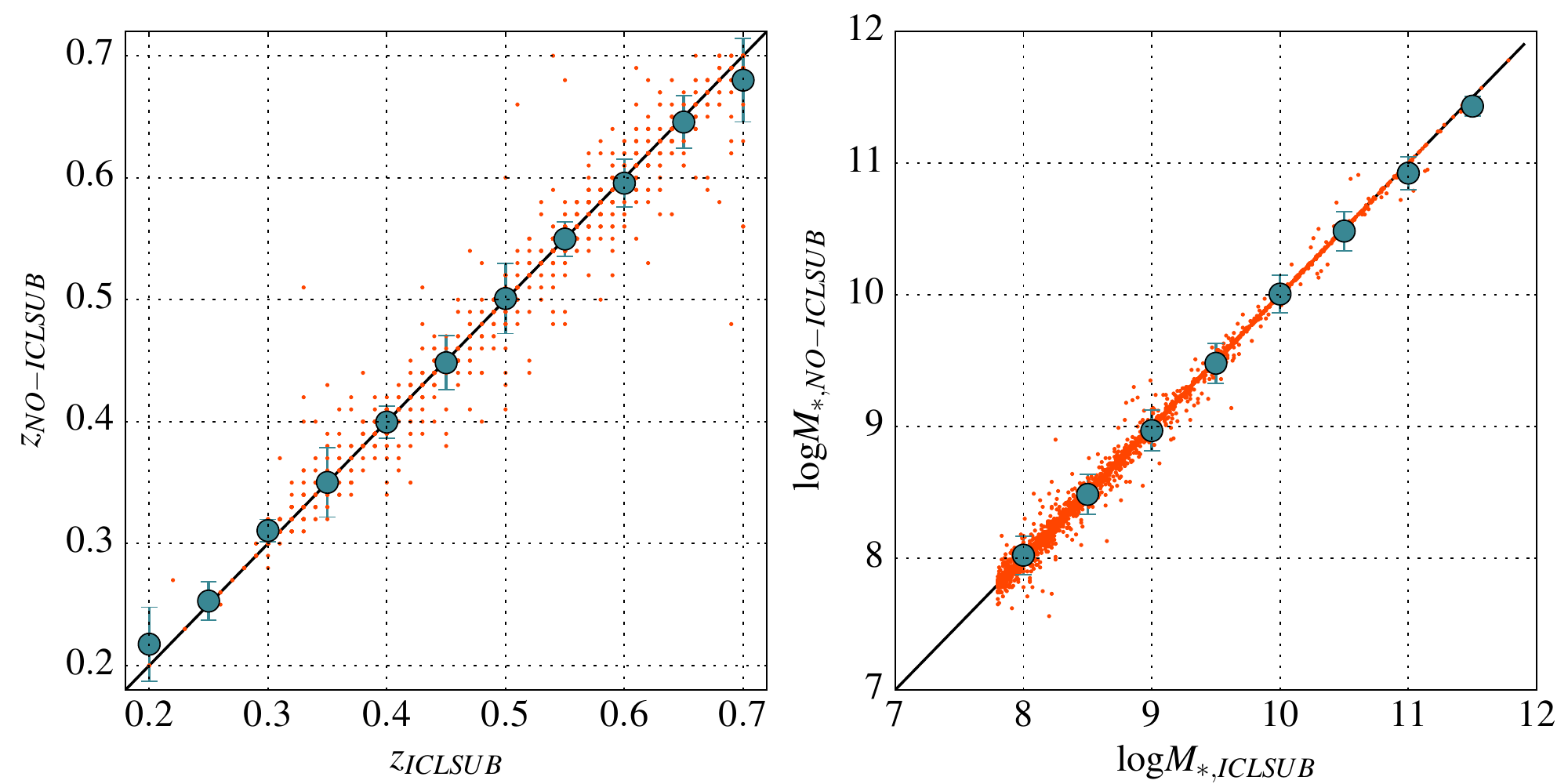}
	\caption{
	Comparisons of photometric redshifts (left) and stellar masses (right) estimated before and after ICL-subtraction.
	Median and normalized median absolute deviation for each redshift and stellar mass bin are shown (blue points with error bars).
	}
\label{fig:delSED}
\end{figure*}

Each CLS pointing contains a substantial amount of diffuse intra-cluster light (ICL). Although \photz\ and stellar mass estimates are are not significantly affected by its presence (Figure \ref{fig:delSED}; scatters are $<0.1\,{\rm dex}$), it is brightest in F160W, which we use for our primary structural measurements. Hence, to minimize possible biases in galaxy parameters, we opt to work with ICL-subtracted images.

Here, we outline the method we developed to consistently subtract a non-parametric ICL component from all CLS images. The procedure makes use of a first round of GALFIT fitting results (Section \ref{sec:result1}) as follows: 
\begin{enumerate}
\item Single Sersic fitting: we first fit each galaxy in a $300\times300\,{\rm pixel}$ postage stamp ($>80$ kpc in our redshift range). We assume a single S\'ersic profile and constant sky background (see Section \ref{sec:result1}). 
\item Reconstruction: using the best fit sky backgrounds, we reconstruct an ICL-only (i.e., background-only) image. 
The value for each pixel in the map is the GALFIT background estimate at that location.
For pixels with multiple postage stamps overlap, the median is calculated by weighting GALFIT's returned $\chi^2/\nu$.
\end{enumerate}
We perform this procedure consistently across all HFF bands (Section \ref{sec:basicData}).

We stress that, in contrast to previous studies \citep[][]{castellano16, livermore16}, we assume nothing about the ICL profile, but only that single S\'ersic profiles well describe {\it non}-ICL components (i.e., individual galaxies).

Some background fluctuations arise from non-ICL components, e.g., zodiacal light in F105W \citep{brammer14}, but these are small compared to the ICL.
Further details of the method are discussed in \citet{morishita16c}.\\

\section*{\small Appendix B\\Photo-$z$ Priors}
\label{sec:PhotoPrior}

When deriving photometric redshifts with EAZY, we multiply the output likelihood distributions by a prior to obtain a best posterior solution.
As mentioned in Section \ref{sec:pzPriors}, we apply one of two different priors for a given galaxy depending on which pointing it is drawn from.

\begin{figure*}
\centering
	\includegraphics[width=8.2cm,bb=0 0 288 288]{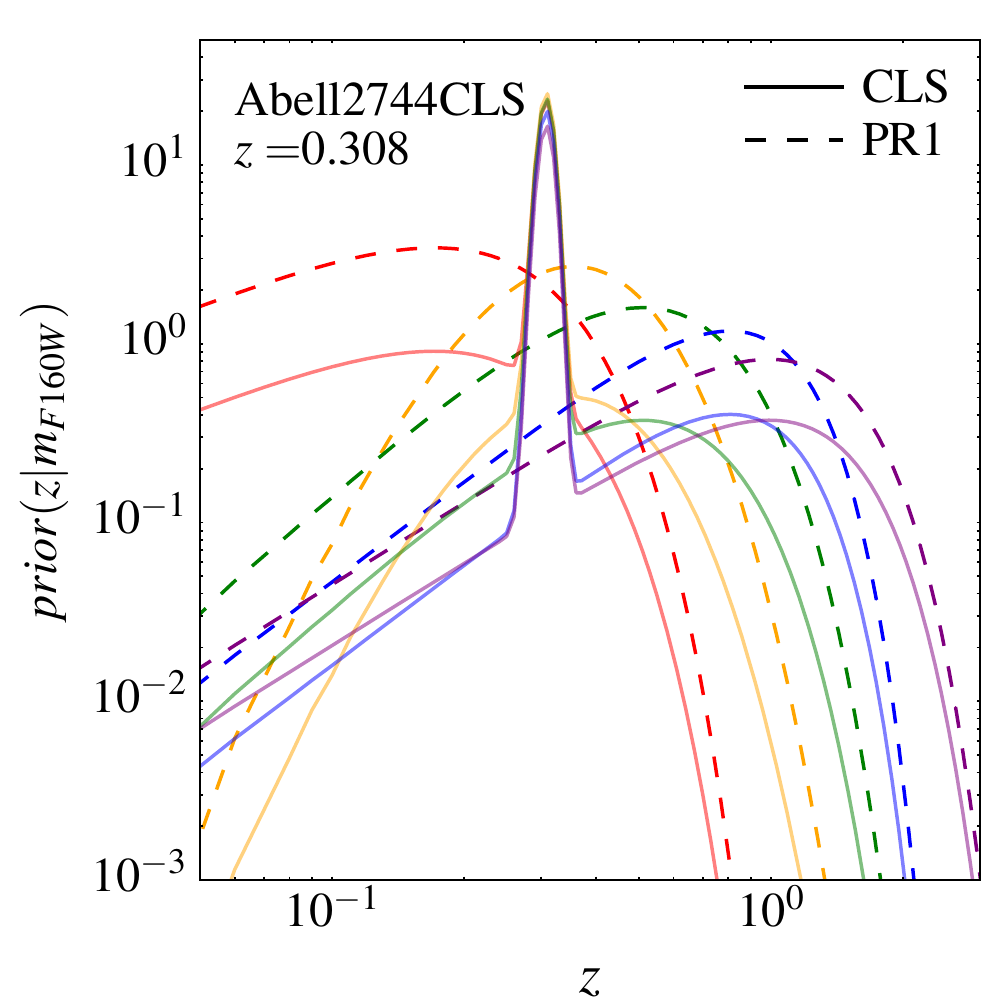}
	\includegraphics[width=8.2cm,bb=0 0 288 288]{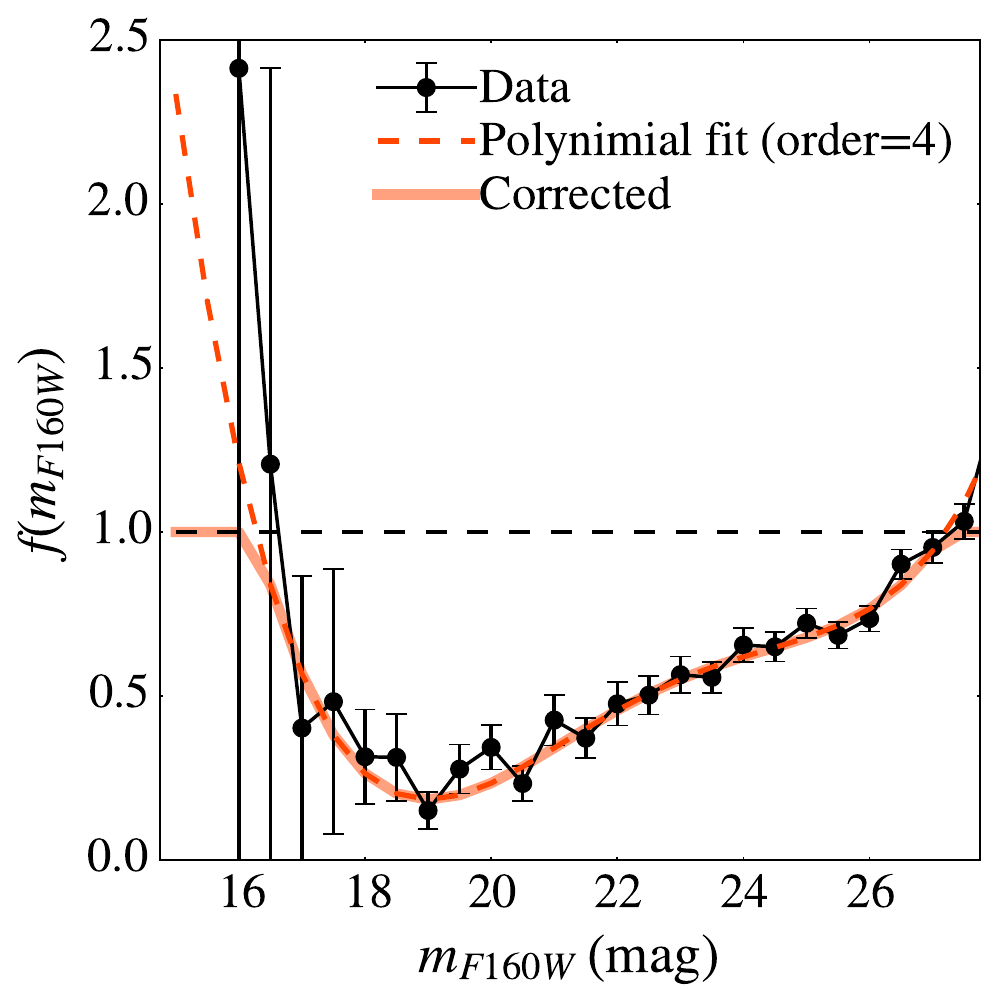}
	\caption{
	Left: photometric redshift priors for PR1 (dashed lines) and CLS (solid lines) provided to EAZY.
	The CLS prior is for Abell2744, where we see strong peak at cluster redshift, $z=0.308$.
	Red to purple lines correspond to $m_{F160W}=19\,{\rm mag}$ to $23\,{\rm mag}$.
	Middle: fraction of the galaxies in PR1 over those in CLS in each magnitude bin (black points).
	The error is calculated by assuming binomial distribution.
	The best fit with a four-ordered polynomial fit is shown with dashed red line.
	The fit is then corrected not to exceed unity (horizontal dashed black line) and shown with solid red line. 
	}
\label{fig:prior}
\end{figure*}

For PR1 sources---mainly field galaxies---we adopt the baseline EAZY prior derived from TAO lightcones for the F160W-band.
The dashed lines in Figure \ref{fig:prior}, left, show these priors, which basically impose the reasonable assumption that apparently brighter objects are more likely to lie at lower redshifts. 

For CLS sources, we modify the above to include the fact that we {\it know} a cluster lies along the line of sight. Because a large fraction of galaxies in CLS pointings will therefore be cluster members, we add to the the PR1 prior a Gaussian centered at the known cluster redshift of width $\sim2000\, \kms$, roughly double the cluster velocity dispersion. We weight the contribution of the field and Gaussian prior components by $f$, the fraction of objets PR1 over those in CLS in bins of apparent magnitude (see main text Equation \ref{eq:prior}).
For more details about using priors in EAZY, see \citet{Brammer08,brammer11}.

The assignment of weights is shown in the right panel of Figure \ref{fig:prior}, and the full CLS prior (for Abell 2744; $z_{\rm cls}=0.308$) is shown by the solid lines at left. 
As expected, an excess of galaxies in CLS emerges over $18 \lesssim m_{\rm F160W} \lesssim 24$  ($f<1$), where cluster members dominate the source counts. 

The factor $f$ exceeds unity at the brightest magnitudes, where stars (rather than cluster members) begin to dominate and sample statistics are weak. We cap the weights, however, at the physically motivated ceiling of $f=1$, denoting that 100\% of sources at that magnitude are {\it not} cluster members.

The weighting factor $f$ is well described by the following analytic form: 
\begin{equation}
	f(x)= {\rm min}[1,~238.8-42.8x+2.9x^2-0.08x^3+0.0009x^4],
\label{eq:CLSpriorWeight}
\end{equation}
where $x=m_{\rm F160W}$. This relation is shown by the red line in Figure \ref{fig:prior}, right. 

\begin{figure}
\centering
	\includegraphics[width=0.47\textwidth]{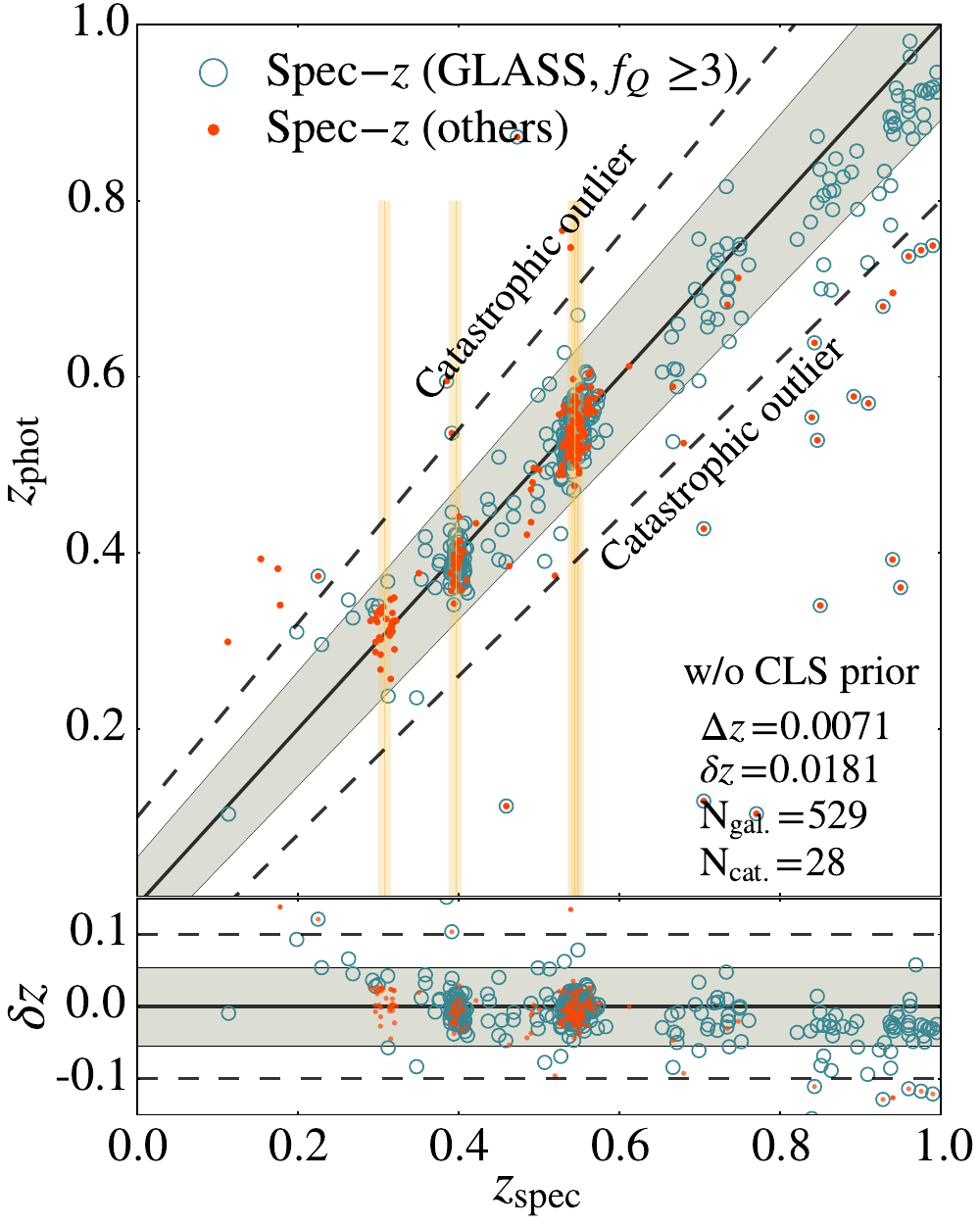}
	\caption{
	Same as Figure~\ref{fig:photz} in the main text, but the phot-$z$ is derived without the cluster prior.
	We see the photometric redshift is more widely scattered at the cluster redshifts, as is also represented in the redshift scatter, $\delta z = 0.0181$.
	We also note that scatters at other redshifts (e.g., \specz$\sim0.2$ and \specz$>0.7$) still remain in this plot, indicating these scatters are not caused by the cluster prior.
	}
\label{fig:woprior}
\end{figure}

Finally, Figure \ref{fig:woprior} shows the same \photz--\specz\ comparison as in Figure \ref{fig:photz}, but without CLS prior.
Here, we see the photometric redshifts are more widely scattered at the known cluster redshifts, with $\delta z = 0.0181$ compared to $\delta z = 0.0073$ obtained using the prior (Section \ref{sec:pzPriors}).\\

\section*{\small Appendix C\\Details on Galaxy Structure Fitting}
\label{sec:Morph}

Galaxy structural parameters are derived with GALFIT \citep{peng10} as in \citet{morishita14}.
Although we only discuss F160W-band half-light radii ($r_{e}$) and S\'ersic indices ($n$) in this study, we perform independent fits in all HFF filter bands so as to be able to characterize (and subtract) the ICL (see Appendix \ref{sec:Aa}).

In the CLS pointing, we re-estimate the structural parameters after subtracting the ICL, while in PR1 we adopt the first fitting results.
This reduces the contamination by ICL, which varies from source to source.
Our fitting procedure is as follows:
\begin{enumerate}
	\item A PSF is empirically constructed using on-image stars with nearby sources masked and local backgrounds correctly subtracted. The TinyTim model is not used because it deviates noticeably from the observed PSF and would lead to underestimated effective radii \citep[][]{morishita14}.
	\item A postage stamp of $200 \times 200\, {\rm pixel}$ for PR1 and $300 \times 300\, {\rm pixel}$ for CLS sources is extracted, centered on the galaxy to be fit. This stamp size difference comes from the fact that cluster centers are crowded, making identification and simultaneous fitting of neighboring objects more critical.
	\item An error map is prepared by cutting the same-size stamp from the image RMS map. This is fed to GALFIT as the ``sigma image.'' 
	\item Other neighboring objects within the same stamp are identified and slated for simultaneous fitting if:
			(i) they are brighter than $18\,{\rm mag}$ (BCGs), or
			(ii) they are $1\,{\rm mag}$ brighter than the target galaxy and reside within $100\,{\rm pixel}$ ($\sim40\,{\rm kpc}$ at $z\sim0.5$).
		Other neighbors are masked.
	\item A mask image is constructed based on the SExtractor segmentation map. This is convolved with a gaussian (kernel size is 1.5~pixel), to cover faint outer envelopes. The pre- and post-convolution masks are then added to make the final mask sent to GALFIT. (Some small pixels fall under the original masking threshold after convolution and so must be replaced).
	\item GALFIT is then run using the above mask and error images to fit a single S\'ersic profile to the target galaxy. SExtractor outputs are taken as the initial guesses for the relevant structural parameters (centroid, position angle, ellipticity, half-light radius), with S\'ersic indices initially set to 2.5 (fits are robust to this assumption). We constrain the GALFIT outputs to $|\delta m|<1\,{\rm mag}$, $|\delta x|<3\,{\rm pixel}$, $|\delta y|<3\,{\rm pixel}$, where $\delta$ is defined with respect to the initial SExtractor values, $0.01<r_e/\,{\rm pixel} <150$, and $0.1<n<8$. 
	Fitting is deemed``successful'' when the GALFIT outputs are within (but not at) the above limits, those failed are excluded from analysis (Appendix \ref{sec:magmass}).
	\item We also fit the background value as a free parameter, with initial value set to 0. This component is assumed to be spatially constant across the stamp. As noted before, we use the best fit sky value of each postage stamp for reconstructing the ICL map.
\end{enumerate}

This procedure is reiterated after subtracting the ICL in CLS pointings.

Notably, though the \hst\ pipeline automatically subtracts a sky background, the final PR1 science images do show non-zero sky levels. These are likely due to time-varying zodiacal light. Hence, a background component is needed for both CLS and PR1 pointings, though it is not further analyzed in the latter.\\

\section*{\small Appendix D\\Detection Completeness, Stellar Mass Limit,\\and GALFIT Completeness}
\label{sec:magmass}

\begin{figure*}
\begin{center}
\includegraphics[width=6.2cm,bb=0 0 288 288]{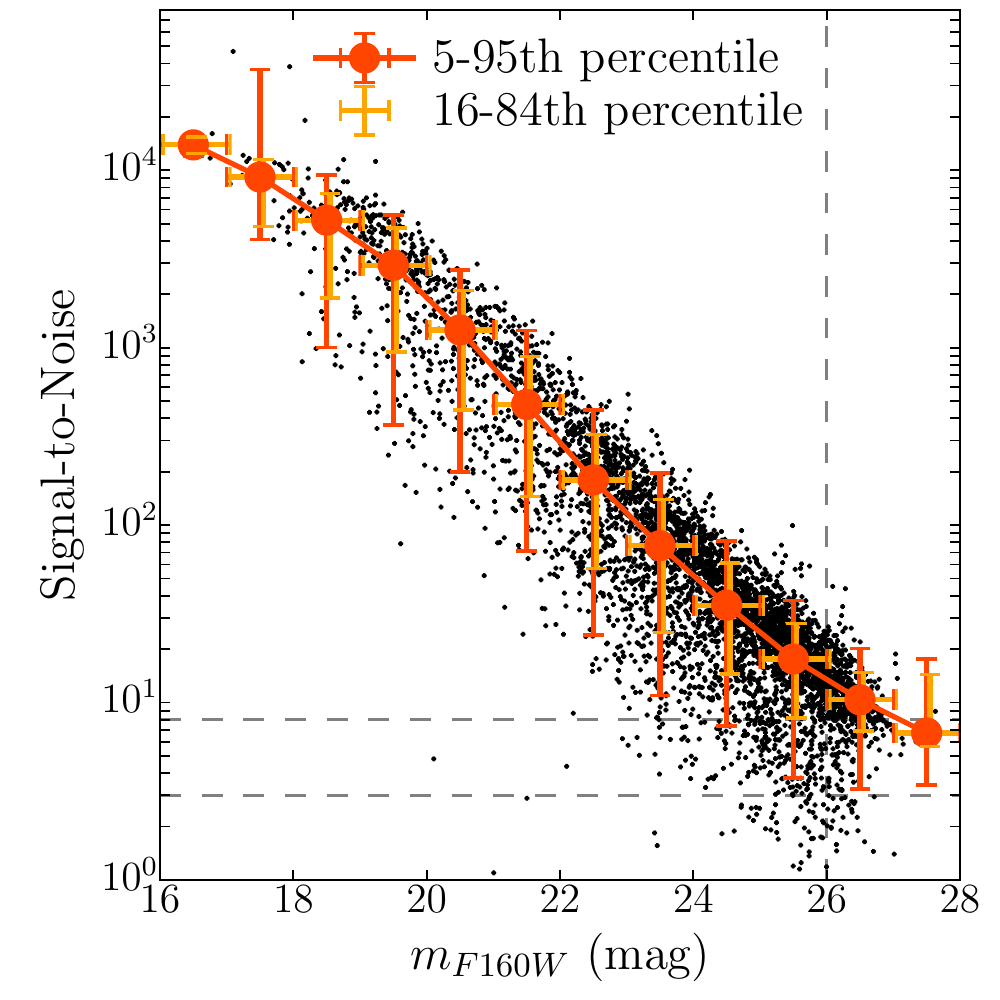}
\includegraphics[width=6.2cm,bb=0 0 288 288]{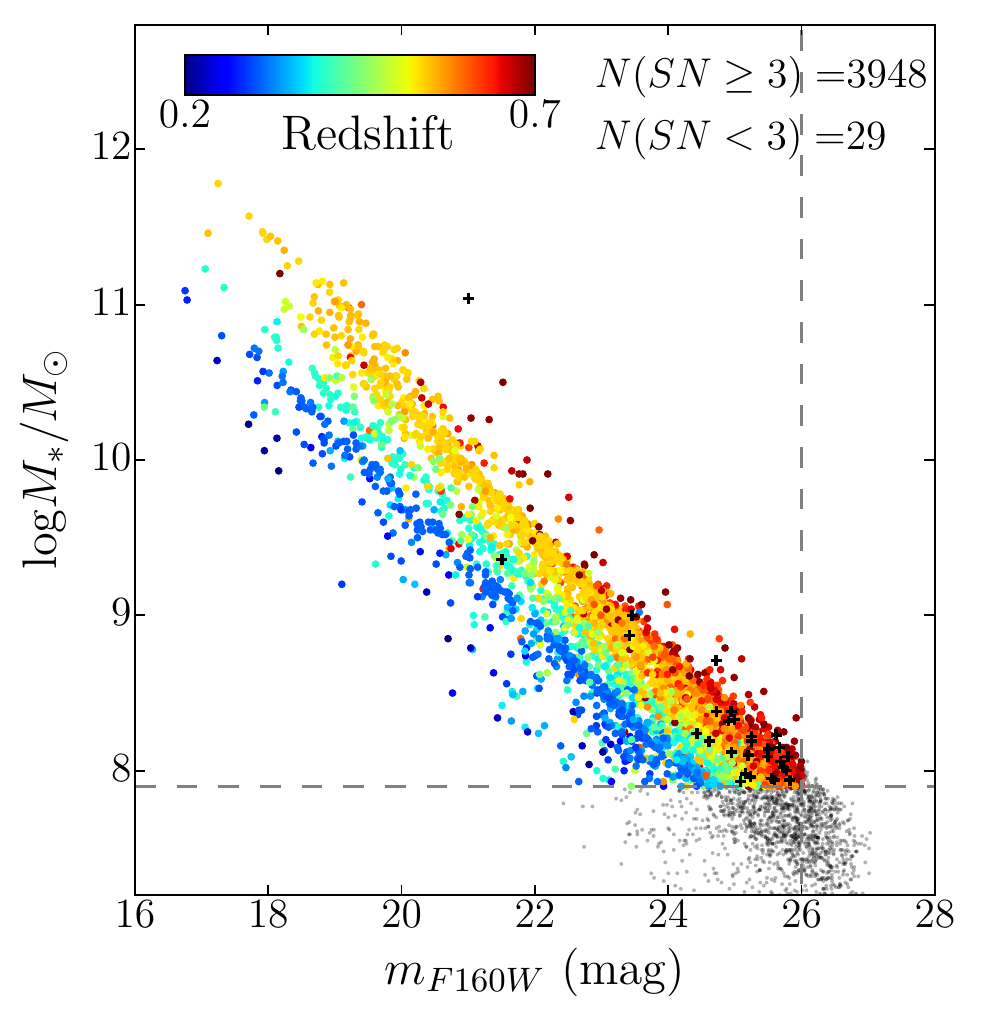}
\caption{
	Left: F160W-band signal-to-noise (S/N) ration versus apparent F160W-band magnitude.
	Median with 5--95th (16--84th) percentile error for each magnitude bin is shown in red (orange) crosses.
	In the faintest magnitude bin ($25<m_{F160W}<26$), more than 95\% of sample are ${\rm S/N}>3$, and $>84\%$ are ${\rm S/N}>8$, where reliable structural parameter can be obtained. 
	Right: apparent magnitude (F160W)-stellar mass plots for our sample.
	Selected samples ($N=3948$) above the magnitude ($m_{F160W}<26.0$) and stellar mass limit ($\logm>7.8$) are color-coded as a function of redshift.
	Excluded samples with $S/N_{F160W}<3$ are shown with black crosses ($N=29$).
	}
\label{fig:magM}
\end{center}
\end{figure*}

Principally, our sample is magnitude-limited to $m_{\rm F160W}<26.0$, reflecting the S/N required to obtain reliable structural parameters. 
Figure~\ref{fig:magM}, left, shows the distribution of our sample in apparent F160W-band magnitude-S/N space.
More than 95\% of sample are ${\rm S/N}>3$, and $>84\%$ are ${\rm S/N}>8$, where others have confirmed that reliable structural parameter can be obtained \citep[e.g.,][]{schmidt14b}.
Cutting our sample at ${\rm S/N}>8$ would increase the success rate in GALFIT (see below), but none of our final results would be significantly affected.

The above flux limit translates into a stellar mass limit based on the FAST results (Section~\ref{ssec:sed}). 
Figure~\ref{fig:magM}, right, shows the relation between $m_{\rm F160W}$, redshift, and inferred stellar mass for our sample. 
At the highest redshifts probed, $m_{\rm F160W} = 26.0\mapsto \logm = 7.8$, hence we adopt this as our final mass completeness limit.

\begin{figure*}
\begin{center}
\includegraphics[width=10.8cm,bb=0 0 504 288]{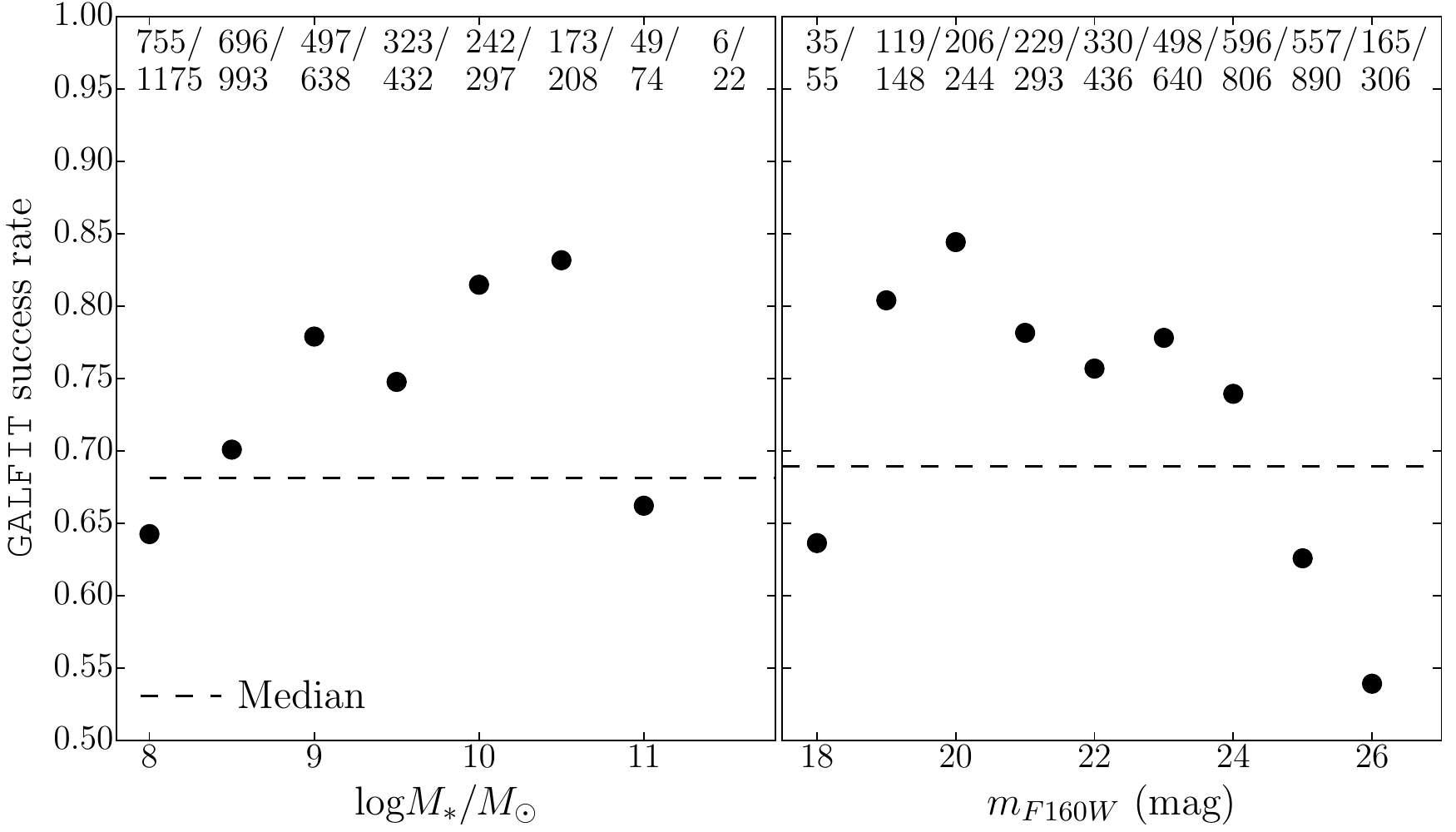}
\caption{
	GALFIT success rate (black points) as a function of stellar mass (left) and F160W-band magnitude (right).
	Median success rate for whole sample, $\approx0.77$, is shown with dashed lines.
	The numbers shown at the top are sample numbers of success and input galaxies at each stellar mass/magnitude bins. 
	}
\label{fig:success}
\end{center}
\end{figure*}

As shown in Figure~\ref{fig:success}, at these stellar mass (left) and magnitude (right), $\sim55\%$ of all sources return successful fits. 
As the {\it detection} limit is much deeper \citep[$m_{F160W}\sim28.7$ in both HFF and XDF;][]{kawamata16, lotz16}, this ensures that all galaxies that can be fit are indeed present in the data, bias in the population of successfully fit sources is minimal.
At these masses, the principal cause of a GALFIT mis-measurement is the presence of a (bright) near-neighbor, a spike in ICL residuals, point sources, or a merger/other violent interaction, as well as faint/noisy objects with $3<{\rm S/N}<8$. 
These are visually identified and excluded from the analysis, though their distribution is shown by the open squares in Figure~\ref{fig:MSR}, and are flagged in the published catalog.
The fitting success rate also drops at the highest masses (Figure \ref{fig:magM}, right).
This is clearly not due to S/N limits, but rather from the fact that such galaxies are large enough that their intrinsic morphological complexity---bars, spiral arms, etc.---starts to matter in the fitting process.
As the total number of these objects is small however, this fact does not quantitatively affect any of our conclusions.

The number of input galaxies are 3948, within our magnitude ($m_{F160W}<26\,{\rm mag}$) and stellar mass/redshift criteria.
Out of the input galaxies, GALFIT converges for 3666.
``Successful'' fits are those whose derived parameters fall within the following limits: 

\begin{enumerate}
	\item Centroids and magnitudes to within $3\,{\rm pixels}$ (in $x$ and $y$) and $1\, {\rm mag}$ of the SExtractor input values, respectively. 
	\item $1<r_e/\,{\rm pixel} <150$, $0.1<n<8$, and $q>0.2$.
\end{enumerate}

After excluding fits that do not meet the above criteria, our final catalog contains 2768 galaxies with robust structural parameters.
This corresponds to a success rate of $\sim67\%$, rising from $\sim64\%$ at $\logm\sim8.0$ to $>80\%$ at $\logm\sim10.0$ (Figure~\ref{fig:success}, left), or $\sim53\%$ at $m_{F160W}\sim26$ to $>80\%$ at $m_{F160W}\sim20$ (Figure~\ref{fig:success}, right).\\

\section*{\small Appendix E\\MCMC}
\label{sec:mcmc}

All best-fit results---for the canonical size-mass relations (3-parameter; Section~\ref{sec:result1}) and the holistic characterization (7-parameter; Section~\ref{sec:result2})---are derived using the EMCEE MCMC code by \citet{foreman13}. This approach enables us to appropriately account for any correlations among the fitting parameters, and also disentangle the intrinsic scatter in the relations from all observational errors.

As an example of our procedure, here we illustrate how to derive our canonical size--mass relation results.

First, set $y=\log r_e/\kpc$ and $x=\logm$. Observed quantities are then represented by,
\begin{equation}
x_i = \bar{x_i} + \epsilon_{x,i}
,\ y_i = \bar{y_i} + \epsilon_{y,i},
\end{equation}
where $\bar{x_i}$ and $\bar{y_i}$ are intrinsic values (which we never know), and $\epsilon_{x, i},\epsilon_{y, i}$ are the (modeled) measurement errors in $x$ and $y$ of $i$th galaxy.
We assume that errors are Gaussian, such that $x_{i}$ and $y_{i}$ can be represented by normal distributions, $N$, with variances $\sigma_{x,i}^2,\sigma_{y,i}^2$. Hence:
\begin{equation}
	x_i = N(\bar{x_i}, \sigma_{x_i}^2 )
	,\ y_i = N(\bar{y_i}, \sigma_{y_i}^2 ).
\end{equation}
We adopt errors from FAST as $\sigma_x$ and from GALFIT as $\sigma_y$.

Now, we can perform the regression.
We adopt a normal linear regression model in literature,
\begin{equation}
\bar{y} = \alpha + \beta (\bar{x}-M_{\rm Norm}) + \sigma,
\end{equation}
where we set $M_{\rm Norm}=9.0$---the median mass of our sample---as the pivot point. This location is chosen so as to minimize covariance in the slope/intercept parameter uncertainties.
The parameters are $\bm{\theta}=(\alpha,\beta,\sigma^2)$, where $\sigma$ is the intrinsic scatter.

We set 20 (50)~random walkers for 3 (7)-parameter fitting, and perform 10,000 MCMC iterations.
Derived parameter estimates and uncertainties are shown in Figures \ref{fig:slope} and \ref{fig:mcmc}, respectively.
It is noted that, though we do not fully sample the covariance of parameters in the full fit, we believe this has little impact on the final parameter values as no strong correlation between any of parameter pairs emerges in in Figure \ref{fig:mcmc}.

\bibliographystyle{apj}

\end{document}